\documentclass[aps,prl,twocolumn,showpacs,amsmath,amssymb,preprintnumbers,10pt,footinbib,superscriptaddress]{revtex4-1}
\usepackage{epsfig}
\usepackage{amsfonts}
\usepackage{amssymb}
\usepackage{enumitem}
\usepackage{amsthm}
\usepackage{subfig}
\usepackage{bm}
\usepackage{hyperref}
\usepackage{mathrsfs}
\usepackage{bbm}
\usepackage{cancel}
\usepackage[dvipsnames]{xcolor}
\usepackage{accents}
\usepackage{relsize}
\usepackage{float, slashed, graphicx, amssymb, amsmath}
\usepackage{shuffle}
\usepackage{pgfplots}
\pgfplotsset{compat=1.14}
\usepackage{tikz}
\usepackage{tikz-3dplot}
\tikzset{
	graviton/.style={line width=.8pt, -latex,decorate, decoration={snake, segment length=4pt,amplitude=1.8pt, pre length=.1cm, post length=.25cm}},
	worldline/.style={gray, line width=1pt},
	worldlineBold/.style={black, line width=.6pt},
	zUndirected/.style={line width=1pt},
	zParticle/.style={line width=1pt,postaction={decorate},decoration={markings,mark=at position .6 with {\arrow[#1]{latex}}}},
	zParticleF/.style={line width=1pt,postaction={decorate}},
	cscalar/.style={line width=1pt,postaction={decorate},decoration={markings,mark=at position .6 with {\arrow[#1]{latex}}}},
	cscalar2/.style={line width=1pt,postaction={decorate},decoration={markings,mark=at position .8 with {\arrow[#1]{latex}}}},
	photon/.style={line width=.8pt, decorate, decoration={snake, segment length=4pt, amplitude=1.8pt,  pre length=.1cm, post length=.1cm}},
    photon2/.style={line width=2pt, gray, decorate, decoration={snake, segment length=5pt, amplitude=1.8pt,  pre length=.1cm, post length=.1cm,}},
    cross/.style={path picture={ 
      \draw[gray,thick]
    (path picture bounding box.south east) -- (path picture bounding box.north west) (path picture bounding box.south west) -- (path picture bounding box.north east);
    }}
}
\usetikzlibrary{hobby,calc,shapes, positioning, backgrounds,trees, decorations, decorations.markings, decorations.pathmorphing, arrows, shapes.geometric, snakes}
\usepackage{tikz-feynman}
\tikzfeynmanset{compat=1.1.0}
\tikzfeynmanset{
graviton/.style={circle, draw=green!60, fill=green!5, very thick, minimum size=7mm}
}
\tikzfeynmanset{
codot/.style={/tikz/shape=circle,
/tikz/fill=red,/tikz/minimum size=0.1cm,/tikz/inner sep=1.8pt}
}
\tikzfeynmanset{sb/.style=
{/tikz/shape=circle,
/tikz/fill=black,
 /tikz/minimum size=0.3cm,/tikz/inner sep=1.pt
 } }
\tikzfeynmanset{myblob/.style=
{/tikz/shape=ellipse,
/tikz/fill=red,
 /tikz/minimum width=0.5cm,
 } }
 \tikzfeynmanset{myblobFF/.style=
{/tikz/shape=circle,
/tikz/fill=black,
 /tikz/minimum width=0.25cm,
 } }
 \tikzfeynmanset{myblob2/.style=
{/tikz/shape=rectangle,
/tikz/fill=red,
 /tikz/minimum width=0.1cm,/tikz/inner sep=1.8pt
 } }
  \tikzfeynmanset{myvertex/.style=
{/tikz/shape=circle,
/tikz/fill=black,
 /tikz/minimum width=0.05cm,
 } }
 \tikzfeynmanset{GR/.style={/tikz/shape=ellipse,/tikz/fill={rgb:black,1;white,2},/tikz/minimum size=0.3cm,} }
\tikzset{box/.pic={\filldraw[fill=black]  (0,0) circle (2.5pt);
				   \filldraw [fill=black] (0.5,0) circle (2.5pt);
			       \draw [line width=5pt] (0,0) -- (0.5,0);}}

\tikzset{wiggle/.style={decorate, decoration=snake}}

 \tikzfeynmanset{HV/.style=
{/tikz/shape=circle,
/tikz/fill={rgb:black,1;white,2},
 /tikz/minimum size=0.01cm,/tikz/inner sep=1.8pt
 } }

\usepackage[T1]{fontenc} % if needed

\def\bell{\boldsymbol{\ell}}
\def\iO{\mathrm{i}0^+}
\newcommand{\I}{\mathrm{i}}

\makeatletter
\newcommand \UPlus {\mathop {\operator@font \uplus }\limits }
\makeatother
\makeatletter
\newcommand \Bigcup {\mathop {\operator@font \bigcup }\limits }
\makeatother
  \def\LabelNote#1{}%\smash{\hbox to\phipt{\raise1ex\hbox{\tiny[#1]}\hss}}}
 \def\Label#1{\label{#1}%
  \smash{\hbox to\phipt{\raise1ex\hbox{\tiny[#1]}\hss}}}

  \newcommand{\eps}{\epsilon}
\def\veps{\varepsilon}

\def\nn{\nonumber}

\newcommand{\black}{\color{black}}
\newcommand{\blue}{\color{blue}}

\def\spa#1.#2{\left\langle#1\,#2\right\rangle}
\def\spb#1.#2{\left[#1\,#2\right]}

\def\be{\begin{equation}}
\def\ee{\end{equation}}
\def\bea{\begin{eqnarray}}
\def\eea{\end{eqnarray}}  
\allowdisplaybreaks

\newcommand{\Li}{\operatorname{Li}_2}
\def\euler{\mathrm{e}}
\def\d{\mathrm{d}}
\newcommand{\apptex}{\textit{Supplemental Material}~\cite{SuppMat}}

\begin{document}

\preprint{HU-EP-26/07-RTG
}

\title{From the Gravitational Compton Amplitude to Black Hole Perturbation Theory 
}
\author{N. Emil J. Bjerrum-Bohr}
\email{bjbohr@nbi.dk}
\author{Gang Chen}
\email{gang.chen@nbi.ku.dk}
\affiliation{Center of Gravity, Niels Bohr Institute, University of Copenhagen, Blegdamsvej 17, DK-2100 Copenhagen, Denmark}
\affiliation{Niels Bohr International Academy, Niels Bohr Institute, University of Copenhagen, Blegdamsvej 17, DK-2100 Copenhagen, Denmark}
\author{Carl Jordan Eriksen}
\email{carl.jordan.eriksen@physik.hu-berlin.de}
\affiliation{Institut f\"ur Physik, Humboldt-Universit\"at zu Berlin,
Zum Gro\ss en Windkanal 2, 10099 Berlin, Germany}
\author{Nabha Shah}
\email{nabha.shah@nbi.ku.dk}
\affiliation{Center of Gravity, Niels Bohr Institute, University of Copenhagen, Blegdamsvej 17, DK-2100 Copenhagen, Denmark}
\affiliation{Niels Bohr International Academy, Niels Bohr Institute, University of Copenhagen, Blegdamsvej 17, DK-2100 Copenhagen, Denmark}

\begin{abstract} 
We employ a single worldline effective field theory in a Schwarzschild--Tangherlini background to compute the gravitational Compton amplitude up to third post-Minkowskian order. After excising the infrared divergences associated with the Weinberg phase, we establish a direct and exact map to results in black hole perturbation theory with a natural resummation and regularization of the forward divergences of the Compton amplitude. We also outline possible applications of this useful computational bridge for Compton amplitudes that include spin and tidal effects.
\end{abstract}  


\maketitle

\section{Introduction}
The ability to detect gravitational waves directly using Earth-based interferometers, along with advances in computational technology for general relativity arising from quantum field theory~\cite{Iwasaki:1971iy,Goldberger:2004jt,Neill:2013wsa,Bjerrum-Bohr:2013bxa,Bjerrum-Bohr:2018xdl,Cheung:2018wkq,Kosower:2018adc}, has led to a renewed theoretical emphasis on deriving analytic results in classical gravity. Several new insights have emerged from studies of the inspiral dynamics \cite{Kalin:2020mvi,Mogull:2020sak, Cristofoli:2019neg,Bern:2019nnu,Antonelli:2019ytb,Bern:2019crd,Parra-Martinez:2020dzs,DiVecchia:2020ymx,Damour:2020tta,Kalin:2020fhe,Bern:2020buy,DiVecchia:2021ndb,DiVecchia:2021bdo,Herrmann:2021tct,Bjerrum-Bohr:2021vuf,Bjerrum-Bohr:2021din,Damgaard:2021ipf,Brandhuber:2021eyq,Bjerrum-Bohr:2021wwt,Bern:2021yeh,Dlapa:2021npj,Dlapa:2021vgp,Jakobsen:2021zvh,Jakobsen:2021lvp,Bern:2021dqo,Bjerrum-Bohr:2022blt,Bjerrum-Bohr:2022ows,Bern:2022jvn,Dlapa:2022lmu,Jakobsen:2022fcj,Adamo:2022ooq,Bern:2022kto,Luna:2023uwd,Damgaard:2023ttc,DiVecchia:2023frv,Heissenberg:2023uvo,Jakobsen:2023ndj,Bjerrum-Bohr:2024hul,Driesse:2024xad,Chen:2024mmm,Bohnenblust:2024hkw,Driesse:2024feo,Chen:2024bpf,Akpinar:2025bkt,Bjerrum-Bohr:2025lpw,Foffa:2019hrb,Blumlein:2019zku,Foffa:2019yfl,Blumlein:2020pog,Blumlein:2020znm, Kalin:2019inp, Jakobsen:2023hig,Kalin:2019rwq,Driesse:2026qiz,Bern:2025wyd,Brandhuber:2025igz,Kim:2025gis} and radiation effects \cite{Kosower:2018adc,Brandhuber:2023hhy,Herderschee:2023fxh,Elkhidir:2023dco,Georgoudis:2023lgf,Caron-Huot:2023vxl,Bohnenblust:2023qmy,Bini:2023fiz,Georgoudis:2023eke,Brandhuber:2023hhl,DeAngelis:2023lvf,Adamo:2024oxy,Bini:2024rsy,Alessio:2024wmz,Georgoudis:2024pdz,Brunello:2024ibk,Brandhuber:2024qdn,Bohnenblust:2025gir,Brunello:2025eso} of two-body mergers. The gravitational Compton amplitude describing graviton scattering off a massive particle is important in this context, serving to probe and characterize stable, curved spacetimes within an effective field theory approach, and is vital for computing results relevant to massive binary scattering via amplitude techniques (see, for instance, \cite{Brandhuber:2021kpo,Brandhuber:2021bsf,
Chen:2022kpm, Aoude:2020onz,
Chung:2020rrz,Chiodaroli:2021eug,Bautista:2021wfy,Aoude:2022trd,Bjerrum-Bohr:2023jau,Chen:2022clh,Bjerrum-Bohr:2023iey,Chen:2024mlx,Cangemi:2023bpe,Bjerrum-Bohr:2024fbt,Bautista:2024emt,Correia:2024jgr,Vazquez-Holm:2025ztz,Ivanov:2022qqt,Saketh:2024juq,Caron-Huot:2025tlq}). 

The Compton amplitude at the second post-Minkowskian order has been studied previously in refs.~\cite{Akpinar:2025huz,Bjerrum-Bohr:2025bqg,Akpinar:2025byi}. The full result for a nonspinning mass was first derived by the present authors~\cite{Bjerrum-Bohr:2025bqg} using worldline effective field theory~\cite{Goldberger:2004jt,Porto:2005ac,Kalin:2020mvi,Mogull:2020sak} in flat and curved backgrounds~\cite{Kosmopoulos:2023bwc,Cheung:2023lnj,Cheung:2024byb}. In this letter, we extend the approach of ref.~\cite{Bjerrum-Bohr:2025bqg} to the third post-Minkowskian order. The infrared structure of the amplitude displays the expected form of the Weinberg phase~\cite{Weinberg:1965nx}, allowing a straightforward extraction of the infrared finite amplitude and the physical differential cross section.

The complete description of gravitational waves scattering against compact objects like black holes, as dictated by general relativity, comes from black hole perturbation theory~\cite{Regge:1957td,Zerilli:1970se,Teukolsky:1972my,Teukolsky:1973ha} and matching to these results systematically identifies the right point particle effective field theory for modeling them. Playing a central role in analyzing gravitational waves in the ringdown regime of a black hole merger process, black hole perturbation theory studies are most naturally conducted in a partial-wave expansion that encodes strong-field dynamics. Establishing an explicit connection between plane-wave amplitudes from post-Minkowskian perturbation theory and partial-wave amplitudes from black hole perturbation theory is therefore essential for extracting and classifying the effects from the finite size of these objects~\cite{Ivanov:2022qqt,Ivanov:2024sds,Bautista:2024emt,Bautista:2024agp,Caron-Huot:2025tlq,Bjerrum-Bohr:2025lpw,Kosmopoulos:2025rfj}. A comparison between these two representations---requiring a projection integral from plane-wave to partial-wave amplitudes---  has been achieved at tree level~\cite{Bautista:2021wfy,Bautista:2022wjf} but an obstacle when going to higher post-Minkowksian orders is the increasingly complex singular integrals arising from forward-limit divergences. \black In this letter, we overcome the challenge by employing a resummation of the leading forward divergences into a compact Newtonian-like contribution~\cite{Dolan:2007ut} that naturally regularizes the singularity. Using it, we establish a direct map between plane-wave and partial-wave amplitudes. We find complete agreement between the scattering amplitudes up to third post-Minkowskian order and therefore obtain the same differential cross section up to next-to-next-to-leading order.   

We envision many interesting applications of our results. Perturbative amplitude relations arising from the non-perturbative isospectrality relations in black hole perturbation theory \cite{Chandrasekhar:1983} could, for instance, lead to new insight on how to resum perturbative amplitude expressions. We also believe that computations of amplitudes with finite-size effects, such as spin and tidal contributions, could receive useful input from the direct link.

The structure of this letter is as follows. We begin by reviewing the computational formalism and discussing the general structure of the classical Compton amplitude. Then, we outline the structure of the third-order post-Minkowskian amplitude, identifying the contributing diagrams and specifying the integral basis required to determine the integrand. The results for the third-order post-Minkowskian amplitude and next-to-leading order cross section are presented next. This is followed by a section describing the comparison of our post-Minkowskian results with black hole perturbation theory by projecting the gravitational post-Minkowskian Compton amplitude to its counterpart in black hole perturbation theory, given in terms of functions in a spin-weighted spherical harmonic basis. Finally, we present our conclusions. 

Throughout, we adopt the mostly-minus metric signature and set $\hbar = c = 1$. 

\section{Gravitational Compton Amplitude }\label{sec:WQFT}%

We start with describing the gravitational interactions of a non-spinning compact object of mass $M$ in an effective field theory approach using the gauge-fixed Einstein--Hilbert action and a minimally coupled worldline action,
\begin{equation}
\label{eq:action}
    S = S_\text{G}[g] + S_\text{wl}[g,x],
\end{equation}
where
\begin{equation}
\begin{aligned}
    &S_\text{G}[g] = -\frac{2}{\kappa_d^2}\int\mathrm{d}^dx\,\sqrt{|g|}\Big(R - \bar g^{\mu\nu}G_\mu G_\nu\Big),\\
    &S_\text{wl}[g,x] = -\frac{M}{2}\int\mathrm{d}\tau\,\big(g_{\mu\nu}\dot x^\mu\dot x^\nu + 1\big).
\end{aligned}
\end{equation}
The gauge-fixing function is $G_\mu = \bar g^{\nu\lambda}\bar\nabla_\nu h_{\lambda\mu} - \textstyle\frac12\bar g^{\rho\sigma}\bar\nabla_\mu h_{\rho\sigma}$ and the bars denote a fixed background metric and its associated covariant derivatives. To regulate formal infinities that arise in loop level computations, we work in $d=4-2\eps$ dimensions. The gravitational coupling constant is $\kappa_d^2 \equiv 32\pi G\tilde\mu^{2\eps}$, with $G$ being the four-dimensional Newton's constant, and $\tilde\mu^2 \equiv (e^{\gamma}/4\pi)\mu^2$. Here, $\mu$ is an arbitrary mass scale coming from dimensional regularization, and $\gamma$ is the Euler--Mascheroni constant.

The object of our study, the scattering of a graviton off a compact object modeled as a point particle, is described by the gravitational Compton amplitude whose relevant kinematics can be summarized as
\begin{equation}
 \begin{tikzpicture}[scale=0.6, baseline=0.1cm]
        \coordinate (v) at (0,0);
        \coordinate (inA) at (-1.8,-.6);
        \coordinate (outA) at (-0.4,-0.2);
        \coordinate (inB) at (0.4,-0.2);
        \coordinate (outB) at (1.8,-.6);
        \coordinate (in) at (-1,-1);
        \coordinate (out) at (1,1);
        \draw [solid] (in) to[out=45-10,in=180+45+20] (v);
        \shade[left color=gray, right color=black, shading angle=70] (v) circle (0.1);
        \foreach \r in {.4,.8,1.2,1.6} {
            \draw[gray!60, thick, rotate=-15] (v) ellipse [x radius={\r*0.9cm}, y radius={\r*0.5cm}];
        }
        \draw [solid,->] (v) to[out=45+20,in=180+45-10] (out);
        \draw[photon,->] (inA) -- (outA);
        \draw[photon,->] (inB) -- (outB);
        \node at (inA) [left] {$p_1^\mu$};
        \node at (outB) [right] {$p_2^\mu$};
        \node at (in) [left] {$v^\mu$};
    \end{tikzpicture}\;\Longrightarrow\hspace{.1cm}\;\begin{tikzpicture}[baseline=(anchor)]
        \coordinate (in) at (-1,0);
        \coordinate (out) at (1,0);
        \coordinate (gin) at (-1,-.93);
        \coordinate (gout) at (1,-.93);
        \coordinate (anchor) at (0,-.93/2);
        \draw [dotted, thick] (in) -- (out);
        \draw [photon] (gin) -- (-.25,-.45);
        \draw [photon] (gout) -- (.25,-.45);
        \draw[black,fill=gray] (0,-.25) circle (.4);
        \node at (gin) [below] {$\veps_1^\mu \veps_1^\nu$};
        \node at (gout) [below] {$\veps_2^{*\rho} \veps_2^{*\sigma}$};
        \node at (in) [left] {$v^\mu$};
    \end{tikzpicture}.
\end{equation}
Here, $p_i^\mu$ and $\veps_i^\mu\veps_i^\nu$ are the momenta and polarization of the gravitons, and $v^\mu$ denotes the velocity of the compact object. The frequency of the external gravitons is defined by $\omega = p_1 \cdot v = p_2 \cdot v$, and the momentum transferred in the process is $q^\mu = p_2^\mu - p_1^\mu$. The magnitude of $q^{\mu}$ is related to the spatial angle $\theta$ between the two graviton momenta and is encoded in
%
%\begin{equation}\label{eq:defx}
 $   x = \frac{|q|}{ 2\omega} = \sin(\frac{\theta}{2})\, , \ 0\leq x\leq 1 \, .$
%\end{equation}
%
The Compton amplitude, which has a non-trivial functional dependence on $x$, can be systematically expanded in a post-Minkowskian series, $n$PM signifying the $n$th order term, and in a series in the dimensional regulator $\eps$ so that
\begin{align}\label{eq:MPMexpand}
	\mathcal{M}=\mathcal{M}_{0}^{\rm 1 PM}+ \sum_{n=2}^{\infty}\sum_{j=1-n}^{\infty}{\epsilon^{j}}\mathcal{M}_{j}^{n\rm PM},
\end{align}
where terms with $j<0$ represent infrared divergences. From Weinberg's soft graviton theorem~\cite{Weinberg:1965nx}, the infrared divergences may be extracted by factoring out the Weinberg phase from the amplitude~\cite{JordanEriksen:2026lox},
   \begin{align}\label{eq:WST}
  	&\mathcal{M}_{\rm finite}(\mu,\veps_1,\veps^*_2)= e^{\I \bar\epsilon / \epsilon}\,\mathcal{M}\big|_{\epsilon\rightarrow 0} \, \nn\\
  	&=\mathcal{M}_{0}^{\rm 1PM}+\mathcal{M}_{0}^{\rm 2PM}+\big(\mathcal{M}_{0}^{\rm 3PM}+\I\bar\epsilon \mathcal{M}_{1}^{\rm 2PM}\big)+\cdots ,
  \end{align}
  where $\bar\epsilon=2GM\omega$ is the post-Minkowskian expansion parameter. Note at third post-Minkowskian order that $\mathcal{M}_{\rm finite}$ has a contribution from the $\epsilon$-linear $\mathcal{M}_{1}^{\rm 2PM}$ \footnote{We thank Gustav Uhre Jakobsen for discussions on this point.}. More generally, at higher orders in the post-Minkowskian expansion, higher powers of $\epsilon$ in the amplitude can also contribute to the finite amplitude and hence contribute to the physical  differential cross section \footnote{We note that the generalized Compton amplitudes contribute to the cross section only at quantum level (see amplitude in \cite{Lee:2021iid} for QED).}, \begin{align}\label{eq:Fin2CS}
  	&{\d\sigma\over\d\Omega}={|\mathcal{M}_{\rm finite}(\mu,\veps_1,\veps_2^{*})|^2\over (8\pi M)^2} \, .  
  	\end{align}
Therefore, positive powers of $\epsilon$ in the amplitude cannot be discarded \emph{a priori}.  We also note that the $\mu$-dependence in the amplitude cancels in the cross section. 

All Feynman rules required for our computation are extracted from the action in eq. \eqref{eq:action} by inserting the background field expansions,
\begin{equation}
    g_{\mu\nu}(x) = \bar g_{\mu\nu}(x) + \kappa_d h_{\mu\nu}(x), \quad x^\mu(\tau) = v^\mu\tau + z^\mu(\tau),
\end{equation}
such that the metric is expanded around the Schwarzschild--Tangherlini solution~\cite{Tangherlini:1963bw} in isotropic coordinates, and the massive particle trajectory around straight-line motion~\cite{Bjerrum-Bohr:2025bqg,Kosmopoulos:2023bwc,Cheung:2023lnj}. Expanding around this solution resums certain classes of Feynman diagrams, reducing the number of diagrams one has to evaluate. To construct the integrand for the third-order Compton amplitude, we add the 13 contributing diagrams \footnote{We use \texttt{xAct} \cite{xAct} to carry out the index contractions in the diagrams.},
\begin{widetext}
\begin{equation}
\begin{aligned}\label{eq:diagrammatic-expansion}
    &\left.\begin{tikzpicture}[baseline=(anchor)]
        \coordinate (in) at (-1,0);
        \coordinate (out) at (1,0);
        \coordinate (gin) at (-1,-.93);
        \coordinate (gout) at (1,-.93);
        \coordinate (anchor) at (0,-.93/2);
        \draw [dotted, thick] (in) -- (out);
        \draw [photon] (gin) -- (-.25,-.45);
        \draw [photon] (gout) -- (.25,-.45);
        \draw[black,fill=gray] (0,-.25) circle (.4);
    \end{tikzpicture} \right\vert_\text{3PM} \;=\; \begin{tikzpicture}[baseline=(anchor),scale=.7]
        \coordinate (anchor) at (0,-.1);
        \coordinate (x) at (0,0);
        \coordinate (gin) at (-1,0);
        \coordinate (x1) at (1,0);
        \coordinate (x2) at (2,0);
        \coordinate (gout) at (3,0);
        \draw [photon] (x) -- (gin) ;
        \draw [photon] (x) -- (x1)-- (x2)-- (gout) ;
        \draw [gray,fill=white,thick] (x) circle (.25) node {\scriptsize${\mathrm{R}}$};
        \draw [gray,fill=white,thick] (x1) circle (.25) node {\scriptsize${\mathrm{R}}$};
        \draw [gray,fill=white,thick] (x2) circle (.25) node {\scriptsize${\mathrm{R}}$};
    \end{tikzpicture} + \begin{tikzpicture}[baseline=(anchor),scale=.7]
    \coordinate (anchor) at (0,-.1);
        \coordinate (x) at (0,0);
        \coordinate (gin) at (-1,0);
        \coordinate (x1) at (1,0);
        \coordinate (x2) at (2,0);
        \coordinate (gout) at (3,0);
        \draw [photon] (x) -- (gin) ;
        \draw [photon] (x) -- (x1)-- (x2)-- (gout) ;
        \draw [gray,fill=white,thick] (x) circle (.25) node {\scriptsize$1$};
        \draw [gray,fill=white,thick] (x1) circle (.25) node {\scriptsize$1$};
        \draw [gray,fill=white,thick] (x2) circle (.25) node {\scriptsize$1$};
    \end{tikzpicture} + \bigg(\begin{tikzpicture}[baseline=(anchor),scale=.7]
    \coordinate (anchor) at (0,-.1);
        \coordinate (x) at (0,0);
        \coordinate (gin) at (-1,0);
        \coordinate (x1) at (1,0);
        \coordinate (x2) at (2,0);
        \coordinate (gout) at (3,0);
        \draw [photon] (x) -- (gin) ;
        \draw [photon] (x) -- (x1)-- (x2)-- (gout) ;
        \draw [gray,fill=white,thick] (x) circle (.25) node {\scriptsize$1$};
        \draw [gray,fill=white,thick] (x1) circle (.25) node {\scriptsize${\mathrm{R}}$};
        \draw [gray,fill=white,thick] (x2) circle (.25) node {\scriptsize${\mathrm{R}}$};
    \end{tikzpicture} + \text{2 perms.}\bigg) \\
    &+ \bigg(\begin{tikzpicture}[baseline=(anchor),scale=.7]
    \coordinate (anchor) at (0,-.1);
        \coordinate (x) at (0,0);
        \coordinate (gin) at (-1,0);
        \coordinate (x1) at (1,0);
        \coordinate (x2) at (2,0);
        \coordinate (gout) at (3,0);
        \draw [photon] (x) -- (gin) ;
        \draw [photon] (x) -- (x1)-- (x2)-- (gout) ;
        \draw [gray,fill=white,thick] (x) circle (.25) node {\scriptsize${\mathrm{R}}$};
        \draw [gray,fill=white,thick] (x1) circle (.25) node {\scriptsize$1$};
        \draw [gray,fill=white,thick] (x2) circle (.25) node {\scriptsize$1$};
    \end{tikzpicture} + \text{2 perms.}\bigg) + \bigg(\begin{tikzpicture}[baseline=(anchor),scale=.7]
    \coordinate (anchor) at (0,-.1);
        \coordinate (x) at (0,0);
        \coordinate (gin) at (-1,0);
        \coordinate (x1) at (1,0);
        \coordinate (gout) at (2,0);
        \draw [photon] (x) -- (gin) ;
        \draw [photon] (x) -- (x1)-- (gout) ;
        \draw [gray,fill=white,thick] (x) circle (.25) node {\scriptsize${\mathrm{R}}$};
        \draw [gray,fill=white,thick] (x1) circle (.25) node {\scriptsize$2$};
    \end{tikzpicture} + \text{1 perm.}\bigg) + \bigg(\begin{tikzpicture}[baseline=(anchor),scale=.7]
    \coordinate (anchor) at (0,-.1);
        \coordinate (x) at (0,0);
        \coordinate (gin) at (-1,0);
        \coordinate (x1) at (1,0);
        \coordinate (gout) at (2,0);
        \draw [photon] (x) -- (gin) ;
        \draw [photon] (x) -- (x1)-- (gout) ;
        \draw [gray,fill=white,thick] (x) circle (.25) node {\scriptsize$1$};
        \draw [gray,fill=white,thick] (x1) circle (.25) node {\scriptsize$2$};
    \end{tikzpicture} + \text{1 perm.}\bigg) + \begin{tikzpicture}[baseline=(anchor),scale=.7]
    \coordinate (anchor) at (0,-.1);
        \coordinate (x) at (0,0);
        \coordinate (gin) at (-1,0);
        \coordinate (gout) at (1,0);
        \draw [photon] (x) -- (gin) ;
        \draw [photon] (x) --  (gout) ;
        \draw [gray,fill=white,thick] (x) circle (.25) node {\scriptsize$3$};
    \end{tikzpicture}, 
\end{aligned}
\end{equation}
\end{widetext}
where we have the deflection and metric insertion subdiagrams,
\begin{align}
\begin{tikzpicture}[baseline=(anchor),scale=.7]
\coordinate (anchor) at (0,-.1);
        \coordinate (in) at (-1,0);
        \coordinate (out) at (1,0);
        \coordinate (x) at (0,0);
        \coordinate (gin) at (-1,0);
        \coordinate (gout) at (1,0);
        \coordinate (v) at (0,0);
        \draw [photon] (x) -- (gin) ;% node [left, below=.5em] {$h_{\mu_1\nu_1}(k_1)$};
        \draw [photon] (x) -- (gout); %node [right, below=.5em] {$h_{\mu_2\nu_2}(k_2)$};
        \draw [gray,fill=white,thick] (x) circle (.25) node {\scriptsize${\mathrm{R}}$};
    \end{tikzpicture}\equiv
\begin{tikzpicture}[baseline=(anchor),scale=.7]
        \coordinate (in) at (-1.3,0);
        \coordinate (out) at (1.3,0);
        \coordinate (x1) at (-.6,0);
        \coordinate (x2) at (.6,0);
        \coordinate (gin) at (-.9,-1);
        \coordinate (gout) at (.9,-1);
        \coordinate (anchor) at (0,-.5);
        \draw [dotted, thick] (in) -- (x1);
        \draw [dotted, thick] (x2) -- (out);
        \draw [zUndirected] (x1) -- (x2) node [midway, below] {$\rightarrow$};
        \draw [photon] (x1) -- (gin) ; %node [left, below=.5em] {$h_{\mu_1\nu_1}(k_1)$};
        \draw [photon] (x2) -- (gout) ; %node [right, below=.5em] {$h_{\mu_2\nu_2}(k_2)$};
        \draw [fill] (x1) circle (.08);
        \draw [fill] (x2) circle (.08);
    \end{tikzpicture}, &&
\begin{tikzpicture}[baseline=(anchor),scale=.7]
        \coordinate (anchor) at (0,-.1);
        \coordinate (in) at (-1,0);
        \coordinate (out) at (1,0);
        \coordinate (x) at (0,0);
        \coordinate (gin) at (-1,0);
        \coordinate (gout) at (1,0);
        \coordinate (v) at (0,0);
        \draw [photon] (x) -- (gin); %node [left, below=.5em] {$h_{\mu_1\nu_1}(k_1)$};
        \draw [photon] (x) -- (gout); % node [right, below=.5em] {$h_{\mu_2\nu_2}(k_2)$};
        \draw [gray,fill=white,thick] (x) circle (.25) node {\scriptsize$n$};
    \end{tikzpicture}\equiv
    \begin{tikzpicture}[baseline=(anchor),scale=.7]
        \coordinate (in) at (-1.,0);
        \coordinate (out) at (1.,0);
        \coordinate (x) at (0,0);
        \coordinate (gin) at (-1.,-1.);
        \coordinate (gout) at (1.,-1.);
        \coordinate (v) at (0,-1.);
        \coordinate (anchor) at (0,-.6);
        \draw [dotted, thick] (in) -- (x);
        \draw [dotted, thick] (x) -- (out);
        \draw [photon2] (x) -- (v) node [midway, right, black] {$\downarrow$};
        \draw [photon] (v) -- (gin) ; %node [left, below=.5em] {$h_{\mu_1\nu_1}(k_1)$};
        \draw [photon] (v) -- (gout);% node [right, below=.5em] {$h_{\mu_2\nu_2}(k_2)$};
        \draw [gray,fill=white,thick] (x) circle (.2) node {\scriptsize$n$};
        \draw [fill] (v) circle (.08);
    \end{tikzpicture}\label{eq:metric-insertion}.
\end{align}
The deflection subdiagram (also known as the recoil vertex \cite{Cheung:2024byb}) represents the process of a graviton being absorbed by the compact object, perturbing its motion, and being reemitted. The $n$th post-Minkowskian metric insertion represents the graviton interacting with the gravitational potential generated by the compact object at that order in the expansion. Due to the non-linearity of general relativity, there is an infinite tower of these metric insertion vertices. We refer to our previous paper for more details \cite{Bjerrum-Bohr:2025bqg}.

The expression for the numerator of the integrand contains scalar products like $\ell_i\cdot\veps_j$. These can be reduced by expanding the polarization vectors $\veps_i^\mu$ in a basis composed of the external kinematics and an additional space-like unit vector $w^\mu$. After tensor reduction, the integrand is spanned by the integral family,
\begin{align}\label{eq:integral-family}
    K_{\vec\nu}^{\vec\lambda} &= \tilde\mu^{4\eps}\int_\ell\frac{\hat\delta^{(\nu_1-1)}(\ell_1\cdot v)\hat\delta^{(\nu_2-1)}(\ell_2\cdot v)}{[\ell_1^2]^{\nu_3}[\ell_2^2]^{\nu_4}[(q-\ell_1-\ell_2)^2]^{\nu_5}}\times \nn\\
    &{[(\ell_1+\ell_2)^2]^{\lambda_1}[(q-\ell_1)^2]^{\lambda_2}(\ell_1\cdot w)^{\lambda_3}(\ell_2\cdot w)^{\lambda_4} \over [(p_1+\ell_1)^2 + i 0^+]^{\nu_6}[(p_1+\ell_1+\ell_2)^2 + i 0^+]^{\nu_7}},
\end{align}
where $\vec\nu = (\nu_1,\nu_2,\nu_3,\nu_4,\nu_5,\nu_6,\nu_7)$, $\vec\lambda = (\lambda_1,\lambda_2,\lambda_{3},\lambda_{4})$. The superscripts on the delta functions signify derivatives and are only needed for the purpose of deriving integration-by-parts relations between the integrals~ \cite{Kotikov:1990kg,Gehrmann:1999as,Smirnov:2008iw,larsen2016integration,Lee:2014ioa}. The first three propagators in the denominator originate entirely from the metric insertion vertices in eq. \eqref{eq:metric-insertion}. They are never on-shell, since they are forced to be spacelike on the whole integration domain. Therefore, their $\iO$ prescription is inconsequential and suppressed in eq.~\eqref{eq:integral-family}. The next two propagators come from the gravitons traversing between any two-point vertices in eq. \eqref{eq:diagrammatic-expansion}, and it is essential to keep track of their $\iO$ prescription. It turns out to be trivial to show that the Feynman and retarded prescriptions are identical for the Compton amplitude. 

With \texttt{LiteRed} \cite{Lee:2012cn,Lee:2013mka}, we reduce the integral family to eight master integrals:~\footnote {Since $K_3 = K_2$ and $K_6 = K_5$, our master integral basis is, in principle, overdetermined.}
\begin{equation}\label{eq:real-masters}
\begin{aligned}
    K_1 &= K_{1,1,0,0,0,1,1}^{0,0,0,0},   
    K_2 = K_{1,1,1,1,0,0,1}^{0,0,0,0}, 
   K_3 = K_{1,1,0,1,1,1,0}^{0,0,0,0} \nn\\ 
    K_4 &= K_{1,1,1,1,1,0,0}^{0,0,0,0},     K_5 = K_{1,1,1,1,1,0,1}^{0,0,0,0},
    K_6 = K_{1,1,1,1,1,1,0}^{0,0,0,0}   \\
    K_7 &= K_{1,1,1,1,1,1,1}^{0,0,0,0},    K_8= K_{1,1,2,1,1,1,1}^{0,0,0,0} \, .
\end{aligned}
\end{equation}
See fig.~\ref{fig:MIs} for a diagrammatic representation. To solve the master integrals, we use the method of differential equations \cite{Kotikov:1990kg,Bern:1993kr,Remiddi:1997ny,Gehrmann:1999as,Henn:2013pwa}. Concretely, we find a canonical basis of master integrals \cite{Henn:2013pwa} which allows us to easily integrate the required integration-by-parts differential equations that appear in the calculation order-by-order in $\eps$. We fix the boundary constants for the three lowest sectors by directly determining the limit of the integrals as $x\to 0$, which is the forward-scattering limit. The boundary constants for the top sector are fixed by imposing the condition that the amplitude be free of all backward scattering singularities as $x\to 1$ and of higher power singularities $x^{-m}$, $m\geq 3$. More details on our procedure and integral results can be found in the \apptex, which includes~\cite{Henn:2014qga,Dlapa:2021qsl,Weinzierl:2022eaz,Baikov:1996cd,Baikov:1996rk,Baikov:2005nv,Frellesvig:2024ymq,Duhr:2019tlz}.

\begin{figure}[t]
    \centering
    \begin{tikzpicture}[scale=.4,transform shape,baseline=(anchor)]
\coordinate(v1)at(-1.3,0);\coordinate(v2)at(1.3,0);\coordinate(v5)at(0,1.15);\coordinate(v6)at(0,.55);
\coordinate(w2)at(3.9,0);\coordinate(w5)at(2.6,1.15);\coordinate(w6)at(2.6,.55);
\coordinate(outL1)at(-1.8,0.75);\coordinate(outL2)at(-1.8,-0.75);
\coordinate(outR1)at(4.4,0.75);\coordinate(outR2)at(4.4,-0.75);
\coordinate(anchor)at(1.3,.05);
\draw[zUndirected,looseness=1.26](v1)to[in=240,out=-60](v2);
\draw[zUndirected,looseness=1.26](v1)to[in=120,out=60](v2);
\draw[zUndirected,red](v5)--(v6);
\draw[zUndirected,looseness=1.26](v2)to[in=240,out=-60](w2);
\draw[zUndirected,looseness=1.26](v2)to[in=120,out=60](w2);
\draw[zUndirected,red](w5)--(w6);
\draw[zUndirected](outL1)--(v1);
\draw[zUndirected](outL2)--(v1);
\draw[zUndirected](w2)--(outR1);
\draw[zUndirected](w2)--(outR2);
\end{tikzpicture}
\begin{tikzpicture}[scale=.4,transform shape,baseline=(anchor)]
\coordinate (out1) at (-1.3,0);
\coordinate (out2) at (2.9,0);
\coordinate (p) at (.8,-2.1);
\coordinate (out3) at (.3,-2.6);
\coordinate (out4) at (2.9,-1.0);
\coordinate (in1) at (-.8,-.5);
\coordinate (in2) at (.8,-.5);
\coordinate (in5) at (2.4,-.5);
\coordinate (v1) at (0,-.2);
\coordinate (v2) at (0,-.8);
\coordinate (v3) at (1.6,-.2);
\coordinate (v4) at (1.6,-.8);
\coordinate (anchor) at (.8,-1.3);
\draw[zUndirected] (out1) -- (in1);
\draw[zUndirected] (out2) -- (in5);
\draw[zUndirected] (out3) -- (p);
\draw[zUndirected] (out4) -- (in5);
\draw[zUndirected] (in1) -- (in2);
\draw[zUndirected] (in2) -- (in5);
\draw[zUndirected] (in1) -- (p);
\draw[zUndirected] (in2) -- (p);
\draw[zUndirected] (in5) -- (p);
\draw[zUndirected,red] (v1) -- (v2);
\draw[zUndirected,red] (v3) -- (v4);
\end{tikzpicture}\,\,
\begin{tikzpicture}[scale=.4,transform shape,baseline=(anchor)]
\coordinate (out1) at (-1.3,0);
\coordinate (out2) at (-1.3,-1.0);
\coordinate (out3) at (2.9,0);
\coordinate (out4) at (1.3,-2.6);
\coordinate (p) at (.8,-2.1);
\coordinate (in1) at (-.8,-.5);
\coordinate (in2) at (.8,-.5);
\coordinate (in5) at (2.4,-.5);
\coordinate (v1) at (0,-.2);
\coordinate (v2) at (0,-.8);
\coordinate (v3) at (1.6,-.2);
\coordinate (v4) at (1.6,-.8);
\coordinate (anchor) at (.8,-1.3);
\draw[zUndirected] (out1) -- (in1);
\draw[zUndirected] (out2) -- (in1);
\draw[zUndirected] (out3) -- (in5);
\draw[zUndirected] (out4) -- (p);
\draw[zUndirected] (in1) -- (in2);
\draw[zUndirected] (in2) -- (in5);
\draw[zUndirected] (in1) -- (p);
\draw[zUndirected] (in2) -- (p);
\draw[zUndirected] (in5) -- (p);
\draw[zUndirected,red] (v1) -- (v2);
\draw[zUndirected,red] (v3) -- (v4);
\end{tikzpicture}
\begin{tikzpicture}[scale=.4,transform shape,baseline=(anchor)]
\coordinate (out1) at (-1.3,0);
\coordinate (out2) at (2.9,0);
\coordinate (p) at (.8,-2.1);
\coordinate (out3) at (.3,-2.6);
\coordinate (out4) at (1.3,-2.6);
\coordinate (in1) at (-.8,-.5);
\coordinate (in2) at (.8,-.5);
\coordinate (in5) at (2.4,-.5);
\coordinate (v1) at (0,-.2);
\coordinate (v2) at (0,-.8);
\coordinate (v3) at (1.6,-.2);
\coordinate (v4) at (1.6,-.8);
\coordinate (anchor) at (.8,-1.3);
\draw[zUndirected] (out1) -- (in1);
\draw[zUndirected] (out2) -- (in5);
\draw[zUndirected] (out3) -- (p);
\draw[zUndirected] (out4) -- (p);
\draw[zUndirected] (in1) -- (in2);
\draw[zUndirected] (in2) -- (in5);
\draw[zUndirected] (in1) -- (p);
\draw[zUndirected] (in2) -- (p);
\draw[zUndirected] (in5) -- (p);
\draw[zUndirected,red] (v1) -- (v2);
\draw[zUndirected,red] (v3) -- (v4);
\end{tikzpicture}\\
\begin{tikzpicture}[scale=.4,transform shape,baseline=(anchor)]
\coordinate (out1) at (-1.3,0);
\coordinate (out2) at (2.9,0);
\coordinate (out3) at (.3,-2.6);
\coordinate (out4) at (2.9,-2.6);
\coordinate (in1) at (-.8,-.5);
\coordinate (in2) at (.8,-.5);
\coordinate (in4) at (.8,-2.1);
\coordinate (in5) at (2.4,-.5);
\coordinate (in6) at (2.4,-2.1);
\coordinate (p) at (.8,-2.1);
\coordinate (v1) at (0,-.2);
\coordinate (v2) at (0,-.8);
\coordinate (v3) at (1.6,-.2);
\coordinate (v4) at (1.6,-.8);
\coordinate (anchor) at (.8,-1.3);
\draw[zUndirected] (out1) -- (in1);
\draw[zUndirected] (out2) -- (in5);
\draw[zUndirected] (out3) -- (p);
\draw[zUndirected] (out4) -- (in6);
\draw[zUndirected] (in1) -- (in2);
\draw[zUndirected] (in2) -- (in4);
\draw[zUndirected] (in1) -- (p);
\draw[zUndirected] (in4) -- (p);
\draw[zUndirected] (in2) -- (in5);
\draw[zUndirected] (in5) -- (in6);
\draw[zUndirected] (in6) -- (in4);
\draw[zUndirected,red] (v1) -- (v2);
\draw[zUndirected,red] (v3) -- (v4);
\end{tikzpicture}\,
\begin{tikzpicture}[scale=.4,transform shape,baseline=(anchor)]
\coordinate (out1) at (2.9,0);
\coordinate (out2) at (-1.3,0);
\coordinate (out3) at (1.3,-2.6);
\coordinate (out4) at (-1.3,-2.6);
\coordinate (in1) at (2.4,-.5);
\coordinate (in2) at (.8,-.5);
\coordinate (in4) at (.8,-2.1);
\coordinate (in5) at (-.8,-.5);
\coordinate (in6) at (-.8,-2.1);
\coordinate (p) at (.8,-2.1);
\coordinate (v1) at (1.6,-.2);
\coordinate (v2) at (1.6,-.8);
\coordinate (v3) at (0,-.2);
\coordinate (v4) at (0,-.8);
\coordinate (anchor) at (.8,-1.3);
\draw[zUndirected] (out1) -- (in1);
\draw[zUndirected] (out2) -- (in5);
\draw[zUndirected] (out3) -- (p);
\draw[zUndirected] (out4) -- (in6);
\draw[zUndirected] (in1) -- (in2);
\draw[zUndirected] (in2) -- (in4);
\draw[zUndirected] (in1) -- (p);
\draw[zUndirected] (in4) -- (p);
\draw[zUndirected] (in2) -- (in5);
\draw[zUndirected] (in5) -- (in6);
\draw[zUndirected] (in6) -- (in4);
\draw[zUndirected,red] (v1) -- (v2);
\draw[zUndirected,red] (v3) -- (v4);
\end{tikzpicture}\,\,
\begin{tikzpicture}[scale=.4,transform shape,baseline=(anchor)]
\coordinate (out1) at (-1.3,0);
\coordinate (out2) at (2.9,0);
\coordinate (out3) at (-1.3,-2.6);
\coordinate (out4) at (2.9,-2.6);
\coordinate (in1) at (-.8,-.5);
\coordinate (in2) at (.8,-.5);
\coordinate (in3) at (-.8,-2.1);
\coordinate (in4) at (.8,-2.1);
\coordinate (in5) at (2.4,-.5);
\coordinate (in6) at (2.4,-2.1);
\coordinate (v1) at (0,-.2);
\coordinate (v2) at (0,-.8);
\coordinate (v3) at (1.6,-.2);
\coordinate (v4) at (1.6,-.8);
\coordinate (anchor) at (.8,-1.3);
\draw[zUndirected] (out1) -- (in1);
\draw[zUndirected] (out2) -- (in5);
\draw[zUndirected] (out3) -- (in3);
\draw[zUndirected] (out4) -- (in6);
\draw[zUndirected] (in1) -- (in2);
\draw[zUndirected] (in2) -- (in4);
\draw[zUndirected] (in3) -- (in4);
\draw[zUndirected] (in3) -- (in1);
\draw[zUndirected] (in2) -- (in5);
\draw[zUndirected] (in5) -- (in6);
\draw[zUndirected] (in6) -- (in4);
\draw[zUndirected,red] (v1) -- (v2);
\draw[zUndirected,red] (v3) -- (v4);
\end{tikzpicture}\,\,
\begin{tikzpicture}[scale=.4,transform shape,baseline=(anchor)]
\coordinate (out1) at (-1.3,0);
\coordinate (out2) at (2.9,0);
\coordinate (out3) at (-1.3,-2.6);
\coordinate (out4) at (2.9,-2.6);
\coordinate (in1) at (-.8,-.5);
\coordinate (in2) at (.8,-.5);
\coordinate (in3) at (-.8,-2.1);
\coordinate (in4) at (.8,-2.1);
\coordinate (in5) at (2.4,-.5);
\coordinate (in6) at (2.4,-2.1);
\coordinate (v1) at (0,-.2);
\coordinate (v2) at (0,-.8);
\coordinate (v3) at (1.6,-.2);
\coordinate (v4) at (1.6,-.8);
\coordinate (anchor) at (.8,-1.3);
\coordinate (m) at (-.8,-1.3);
\draw[zUndirected] (out1) -- (in1);
\draw[zUndirected] (out2) -- (in5);
\draw[zUndirected] (out3) -- (in3);
\draw[zUndirected] (out4) -- (in6);
\draw[zUndirected] (in1) -- (in2);
\draw[zUndirected] (in2) -- (in4);
\draw[zUndirected] (in3) -- (in4);
\draw[zUndirected] (in3) -- (in1);
\draw[zUndirected] (in2) -- (in5);
\draw[zUndirected] (in5) -- (in6);
\draw[zUndirected] (in6) -- (in4);
\draw[zUndirected,red] (v1) -- (v2);
\draw[zUndirected,red] (v3) -- (v4);
\fill (m) circle (4pt);
\end{tikzpicture}
    \caption{Diagrams representing the master integrals ordered left to right from $K_1$ to $K_4$ and then $K_5$ to $K_8$. Red lines denote velocity-cut propagators. The dot on the second double box signifies a squared propagator.}
    \label{fig:MIs}
\end{figure}
We find that the third post-Minkowskian order amplitude up to $\mathcal{O}(\epsilon)$,
\begin{align}
	\mathcal{M}^{\rm 3PM}=   -\frac{\bar\eps^2}{2 \eps^2}\mathcal{M}^{1\rm PM}_{0} -\frac{\I\bar\eps}{\eps}\mathcal{M}^{2 \rm PM}_{0}+\mathcal{M}^{3 \rm PM}_{0}+ \mathcal{O}(\epsilon) \, ,
\end{align}
has the expected infrared behaviour from Weinberg's soft theorem \cite{Weinberg:1965nx}. Then, using eq.~\eqref{eq:WST} and inserting the linear-in-$\epsilon$ term from the second-order post-Minkowskian amplitude, extracted from the $d$-dimensional expression presented in \cite{JordanEriksen:2026lox}, we obtain the finite amplitude at third post-Minkowskian order, expressed here in terms of the transcendental functions appearing in the master integrals,
\begin{align}\label{eq:3PM-result}
  &\mathcal{M}_{\rm finite}^{\rm 3PM}
  =
  \mathcal{M}^{\rm 3PM}_{0}+i\bar\epsilon  \mathcal{M}^{\rm 2PM}_{1}
  =
  \frac{\pi M (2 G M \omega )^3}{\omega}
  \notag\\
  &\!\!\times\!\!\bigg[
    d_1\big(G_+(x)+\I\pi\log(1+x)\big)
    \!+\!
    d_2\Big(G_-(x)+\I\pi\log\frac{4x}{1+x}\Big)
    \notag\\
    &+
    d_3\log^2\frac{4\omega^2x^2}{\mu^2}
    +
    d_4\Big(4\zeta(2)-\log^2\frac{4\omega^2}{\mu^2}\Big)
    +
    d_5\log x^2
    \notag\\
    &+
    (d_6+d_7 \, \I\pi)\log\frac{4\omega^2}{\mu^2}
    +
    d_\text{Re}
    +
    d_\text{Im}
  \bigg],
\end{align}
where we have used
\begin{equation}\label{eq:Gplusminus-definition}
    G_\pm(x) = \Li(x)\pm\Li(-x) + (\log(1-x)\pm\log(1+x))\log x.
\end{equation}
The coefficients, which are given in the \apptex, are all in terms of two gauge-invariant functions,
\begin{equation}\label{eq:gauge-invs}
    \mathsf{F}_1 = \frac{(v\cdot f_1\cdot f_2^*\cdot v)}{\omega^2},\ \
    \mathsf{F}_2 = \frac{(v\cdot f_1\cdot p_2)(v\cdot f_2^*\cdot p_1)}{\omega^4},
\end{equation}  
where $f_{i}^{\mu\nu} = p_{i}^{\mu}\varepsilon_{i}^{\nu}-p_{i}^{\nu}\varepsilon_{i}^{\mu}$.
The poles at $x\to1$ and $x\to1/\sqrt{2}$ are spurious and originate in the tensor and integration-by-parts reduction. We remark  that the leading forward divergences up to third post-Minkowskian order in $\mathcal{M}_{\rm finite}$ go as
 \begin{align}\label{eq:3PMdiv}
 	\text{3PM: }
    x^{-2}\log^2x
    ,
    \quad
    \text{2PM: }
    x^{-2}\log x
    ,
    \quad
    \text{1PM: }
    x^{-2}
    .
 \end{align}

Using eqs.~\eqref{eq:WST} and \eqref{eq:Fin2CS}, together with the first and second order post-Minkowskian amplitude results in \cite{Bjerrum-Bohr:2025bqg}, one can directly get the leading order, next-to-leading order, and next-to-next-to-leading order cross sections from the expression, 
   \begin{align}
  	\bigg({d\sigma\over d\Omega}\bigg)&={\lvert\mathcal{M}_{0}^{\rm 1 PM}\rvert^2+2{\rm Re}(\mathcal{M}_{0}^{\rm 1 PM}\mathcal{M}_{0}^{\rm 2 PM})\over (8\pi M)^2} \nn\\
    &+{\lvert\mathcal{M}_{0}^{\rm 2 PM}\rvert^2+2{\rm Re}(\mathcal{M}_{0}^{\rm 1 PM} \mathcal{M}_{\rm finite}^{\rm 3PM})\over (8\pi M)^2 } +\cdots \,.
  	\end{align}
Working in the lab frame, one can take $\theta_1=0$ and relabel $\theta_2=\theta$ in the following general expressions for the external data present in the amplitude, $ v=(1,0,0,0),\,
    p_j=\omega(1,\sin \theta_j  ,0,\cos \theta_j ),\,
   \varepsilon^{+}_j=\frac{1}{\sqrt{2}}(0,\cos \theta_j    ,
  i  ,-\sin \theta_j ),\,
   \varepsilon^{-}_j=\frac{1}{\sqrt{2}}(0,\cos \theta_j   ,
  -i  ,-\sin \theta_j )$, so that the cross-section can be rephrased as receiving contributions from a helicity-conserving function $f$ and helicity-reversing function $g$,
\begin{align}
	{d\sigma\over d\Omega}=|f_{\rm amp}(\theta)|^2+|g_{\rm amp}(\theta)|^2\, ,
\end{align} 
where $
	f_{\rm amp}(\theta)={1\over 8\pi M}{\cal M}_{\rm finite}(\mu,\veps^+_1,\veps_2^{+*})$ and $
	g_{\rm amp}(\theta)={1\over 8\pi M}{\cal M}_{\rm finite}(\mu,\veps^+_1,\veps_2^{-*})$. 
Using explicit expressions for the amplitudes, the cross-section up to next-to-leading order is  %
  \begin{align}
{d\sigma\over d\Omega}&=(GM)^2\Bigg[\frac{(x^2-1)^4+x^8}{x^4}+\bar{\epsilon}\frac{\pi(15 (x+1)^2-2 x)}{4x^3}\nn\\
&\times (x-1)^4+\bar{\epsilon}^2\Big(
c_1 G_{+}(x)+c_2 G_{-}(x)+c_3+\pi ^2 c_4\nn\\
&+c_5 \log (x)+c_6 \log ^2(x)
\Big)\Bigg],
\end{align}
where the coefficients $c_i$ are provided in the \apptex.  In the plot shown in fig.~\ref{fig:cs3}, we display the leading order, next-to-leading order, and next-to-next-to-leading order cross sections separately as functions of $\theta$ at $\bar\epsilon=0.3$.  The leading order is dominant at all angles and indicates the efficiency of the post-Minkowskian expansion (at $\bar\epsilon\leq 0.3$).
\begin{figure}
\begin{tikzpicture}[scale=0.9]
\begin{axis}[
title={},   axis lines=left,
			xlabel=$\theta$, 
			ylabel= ${\displaystyle d\sigma\over d\Omega}/ (GM)^2$, ylabel style={rotate=0,at={(-0.09,0.7)},anchor=south},
			legend style={at={(0.25,1.)},anchor=north},
			ymajorgrids=true,
			grid style=dashed,
            every axis plot/.append style={thick},enlargelimits=0.05]
\addplot[color=blue, mark=star] coordinates {(81,0.806229) (83.25,0.695406) (85.5,0.610532) (87.75,0.54671) (90,0.5) (92.25,0.467219) (94.5,0.44579) (96.75,0.433618) (99,0.428997) (101.25,0.430533) (103.5,0.437085) (105.75,0.447712) (108,0.461641) (110.25,0.478229) (112.5,0.496943) (114.75,0.517336) (117,0.539032) (119.25,0.561713) (121.5,0.585107) (123.75,0.60898) (126,0.633129) (128.25,0.657374) (130.5,0.681557) (132.75,0.705537) (135,0.729185) (137.25,0.752383) (139.5,0.775023) (141.75,0.797006) (144,0.818237) (146.25,0.83863) (148.5,0.858103) (150.75,0.876579) (153,0.893986) (155.25,0.910257) (157.5,0.92533) (159.75,0.939147) (162,0.951656) (164.25,0.962808) (166.5,0.972561) (168.75,0.980878) (171,0.987726) (173.25,0.99308) (175.5,0.99692)};
\addlegendentry{LO}

\addplot[color=brown, mark=square*,mark size=1pt] coordinates {(81,0.256613) (83.25,0.205423) (85.5,0.164046) (87.75,0.130646) (90,0.103727) (92.25,0.0820763) (94.5,0.0647021) (96.75,0.0507973) (99,0.0397027) (101.25,0.0308805) (103.5,0.023892) (105.75,0.0183793) (108,0.014051) (110.25,0.0106699) (112.5,0.00804357) (114.75,0.00601598) (117,0.00446113) (119.25,0.00327754) (121.5,0.00238377) (123.75,0.00171478) (126,0.00121883) (128.25,0.000855041) (130.5,0.000591274) (132.75,0.000402463) (135,0.000269208) (137.25,0.000176627) (139.5,0.000113418) (141.75,0.0000710971) (144,0.0000433769) (146.25,0.0000256649) (148.5,0.0000146626) (150.75,8.04579e-6) (153,4.21272e-6) (155.25,2.08735e-6) (157.5,9.68366e-7) (159.75,4.1477e-7) (162,1.60936e-7) (164.25,5.5083e-8) (166.5,1.59947e-8) (168.75,3.70927e-9) (171,6.20866e-10) (173.25,6.20446e-11) (175.5,2.41777e-12)};
\addlegendentry{NLO}

\addplot[color=red, mark=triangle*,mark size=1.5pt] coordinates {(81,0.177348) (83.25,0.141845) (85.5,0.112816) (87.75,0.0891326) (90,0.069866) (92.25,0.0542481) (94.5,0.0416444) (96.75,0.0315296) (99,0.0234692) (101.25,0.0171032) (103.5,0.0121335) (105.75,0.00831348) (108,0.00543879) (110.25,0.00334051) (112.5,0.00187893) (114.75,0.000938526) (117,0.000423847) (119.25,0.00025598) (121.5,0.000369658) (123.75,0.000710834) (126,0.00123466) (128.25,0.00190379) (130.5,0.00268701) (132.75,0.00355803) (135,0.0044946) (137.25,0.00547767) (139.5,0.00649077) (141.75,0.00751951) (144,0.00855115) (146.25,0.00957429) (148.5,0.0105786) (150.75,0.0115548) (153,0.0124941) (155.25,0.0133887) (157.5,0.0142311) (159.75,0.0150146) (162,0.0157329) (164.25,0.0163806) (166.5,0.0169524) (168.75,0.0174441) (171,0.0178516) (173.25,0.018172) (175.5,0.0184026)};
\addlegendentry{NNLO}
\end{axis}
  % Axes
  \draw[->] (5,2.5,0) -- (5.9,2.5,0) node[anchor=north east] {$$};
  \draw[->] (5,2.5,0) -- (5,3.4,0) node[anchor=east] {$$};
 % \draw[->] (0,0,0) -- (0,0,-0.8) node[anchor=north] {$y$};
  % p1 direction
  \coordinate (p1) at (5,1.6,0);
 %\node at (0,-0.5,0) {$~$};
  \draw[->, very thick, black] (p1) -- (5,2.5,0) node[below right] {$p_1$};
  % polarization example vectors
  \coordinate (etheta) at (5.6,2.5,0);
  \draw[->, thick, black] (5,2.5,0) -- ($(8,8.8,0)!0.88!(etheta)$) node[right] {$p_2$};
  % angle arc
  \tdplotdefinepoints(5,2.5,0)(5,3.5,0)(5.9,3.4,0)
  \tdplotdrawarc[thin]{(5,2.5,0)}{0.55}{40.5}{90}{anchor=south}{$~\theta$};
\end{tikzpicture}
\caption{Differential cross section at $\bar\epsilon=0.3$ showing leading order (LO), next-to-leading order (NLO), and next-to-next-to-leading order (NNLO) contributions.}
\label{fig:cs3}
\end{figure}

\section{Comparison with black hole perturbation theory}
\label{scattering function}
In black hole perturbation theory, the helicity-conserving  and helicity-reversing contributions to the differential cross section are expressed as an infinite sum involving spin-weighted spherical harmonic functions \cite{Matzner:1977dn,Dolan:2007ut,Dolan:2008kf} written here in a modified form,
\begin{align}\label{eq:fbgb}
f_{\rm B}(\theta)
&=
\frac{\pi e^{-i\Phi}}{i\omega}\sum_{l=2}^{\infty}
{}_{-2}Y_{l2}(0)\,
{}_{-2}Y_{l2}(\theta)\, \Delta^{+}_{l2},\nn\\
g_{\rm B}(\theta)
&=
\frac{\pi e^{-i\Phi}}{i\omega}\sum_{l=2}^{\infty}
 (-1)^{l}
{}_{-2}Y_{l2}(0)\,
{}_{-2}Y_{l2}(\pi-\theta)\Delta^{-}_{l2} ,
\end{align}
where $\Phi\equiv 2\bar\epsilon \ln(2\bar\epsilon)-\bar\epsilon$ and $\Delta^{\pm}_{l2}=e^{2i\delta^{+}_{l2}} \pm  e^{2i\delta^{-}_{l2} }$. We drop the subtraction of $2$ in the original $f_{\rm B}(\theta)$ to make the definition consistent with the post-Minkowskian amplitude and also multiply by an $l$-independent overall phase that does not affect the physical differential cross section. The parity- and $l$-dependent phases are obtained from the Mano-Suzuki-Takasugi solution \cite{Mano:1996vt} of the Teukolsky equation~\cite{Teukolsky:1972my,Teukolsky:1973ha} %\NS{[Regge-Wheeler?]} \red [Carl: Since the phase shift is extracted from the NP scalar $\psi_4$, it is the Teukolsky equation, right?] \black,
\begin{align}\label{eq:expPhase}
	e^{2i\delta_{lm}^{\pm}}&=(-1)^{l+1}{C^{(0)}_{\rm TS}\over 16\omega^4}{B^{\rm (ref)}_{{lm}}\over B^{\rm (inc)}_{{lm}}},
\end{align}
where $C^{(0)}_{\rm TS}=l(l^2-1)(l+2)\pm 6i \bar\eps $ is the Teukolsky-Starobinsky constant in the spinless limit. $B^{\rm (ref)}$ and $B^{\rm (inc)}$ are obtained from the asymptotic behavior of the Mano-Suzuki-Takasugi solution for the Newman-Penrose scalar $\psi_4$ and we refer to \cite{Mano:1996vt,Sasaki:2003xr} for their explicit values~\footnote{We recommend the efficient program described in \cite{BHPToolkit} for calculating all the relevant quantities and also point the reader to \cite{Markovic:2025kvr} for a program to calculate phase shifts using \cite{BHPToolkit}.}. If we define components of the scattering amplitudes under projection to the spin-weighted  spherical harmonic basis as 
\begin{align}
	(f)^{(l)}&\equiv 2\pi\int_{0}^{\pi}  f(\theta) \, {}_{-2}Y_{l2}(\theta)\sin(\theta)d\theta, \nn\\
    (g)^{(l)}&\equiv 2\pi\int_{0}^{\pi}  g(\theta) \, {}_{-2}Y_{l2}(\pi-\theta)\sin(\theta)d\theta ,
\end{align}
it is straightforward to deduce a non-perturbative relation between helicity-conserving and helicity-reversing functions,
\begin{align}\label{eq:NPTRelation}
	g^{(l)}_{\rm B}={6i\bar\epsilon\over (-1)^{l} l(l^2-1)(l+2)}f^{(l)}_{\rm B},
\end{align}
as the parity-dependent phase arises only from the Teukolsky-Starobinsky constant. This relation is also proved in \cite{Futterman_Handler_Matzner_1988} based on Zerilli's radial equation \cite{Zerilli:1970se}, see also \cite{Dolan:2007ut}.

To demonstrate the equivalence between the Compton amplitude in a post-Minkowskian expansion and expressions from black hole perturbation theory, it is useful to project the Compton amplitude onto the same complete basis of spherical harmonic functions.  In order to do so and match known results,  we set the scale parameter in the Compton amplitude to $\mu = 2e^{-\gamma}\omega$, and first proceed to regularize the forward scattering divergences in eq.~\eqref{eq:3PMdiv} that prevent successful projection.  Covariantizing the results of long-range (Newtonian) contributions to gravitational wave scattering in \cite{Andersson:2000tf, Dolan:2007ut}, we find that a resummation of the leading forward divergent components from each post-Minkowskian order suffices as a natural regularization. While this contribution contains terms at all orders in the post-Minkowskian expansion, it arises solely from iterations of the tree-level Newtonian potential. Thus, we arrive at 
\begin{align}\label{eq:SingSub}
\mathcal{M}_{\rm finite}(2e^{-\gamma}\omega,\veps_1,&\veps_2^*)
=32\pi G M^2
\frac{\Gamma (1-i \bar\epsilon )}{ \Gamma (1+i \bar\epsilon )}
\frac{\mathsf{F}_1^2}{4x^{2 (1-i \bar \epsilon )}}\nn\\
+&\sum_{n=2}^\infty 
\mathcal{M}_{\rm finite,rm}^{n\rm PM}( 2e^{-\gamma}\omega,\veps_1,\veps_2^*)\, ,
\end{align}
where the subscript ``rm'' denotes remaining terms that are free of the leading forward divergent contribution. \blue  \black We find (to the order we work) that the leading forward divergent contribution can be understood from considering the resummation of graphs with a double cut at each loop,
\begin{align}
\begin{tikzpicture}[baseline={([yshift=-0.8ex]current bounding box.center)}]\tikzstyle{every node}=[font=\scriptsize]	
\begin{feynman}
    	 \vertex (p1) {};
    	 \vertex [right=1.3cm of p1] (b1)[]{};
    	 \vertex [right=1.3cm of b1] (b2)[]{};
    	  \vertex [right=1.3cm of b2] (b3)[]{};
    	 \vertex [right=1.3cm of b3] (p4){};
    	  \vertex [right=0.45cm of b1] (la)[]{};
    	 \vertex [above=0.5cm of p1](p2){};
    	 \vertex [right=1.3cm of p2] (u1) [HV]{H};
    	 \vertex [right=1.3cm of u1] (u2) [HV]{H};
    	 \vertex [right=1.3cm of u2] (u3)[HV]{H};
    	  \vertex [above=0.5cm of p4](p3){};
    	  \vertex [right=0.65cm of u1] (cutu1) []{$\textcolor{red}{\rule{1.5pt}{2ex}}$};
    	  \vertex [right=0.65cm of u2] (cutu2) []{$\textcolor{red}{\rule{1.5pt}{2ex}}$};
    	  \vertex [below=0.45cm of cutu1] (cutd1) []{$\textcolor{red}{\rule{1.5pt}{2ex}}$};
    	   \vertex [below=0.45cm of cutu2] (cutd2) []{$\textcolor{red}{\rule{1.5pt}{2ex}}$};
    	  \vertex [above right=0.4cm of p4] (tet) []{$\bullet \bullet \bullet$};
    	  \diagram* {
(p1) -- [ultra thick,photon,out=10,in=-130,looseness=0.5] (u1), (u3)-- [ultra thick,photon,out=-55,in=-200,looseness=0.7] (p4),
    	   (u1)-- [ultra thick,photon,out=-75,in=-105,looseness=0.5,min distance=0.2cm] (u2),(u2) -- [ultra thick,photon,out=-75,in=-105,looseness=0.5,min distance=0.2cm] (u3), (p2) -- [ultra thick] (u1)-- [ultra thick] (u2)-- [ultra thick] (u3)-- [ultra thick] (p3),
    	  }; 
    \end{feynman}  
    \end{tikzpicture}
    \end{align}
The logic behind this argument starts from the one-loop amplitude result, where one arrives at contributions with the form $(\log(x)+\gamma){1\over x^2}$. When adding a new rung, these one-loop terms reappear together with new higher-order loop terms. Since the total contribution from summing all loop orders will consist of all lower order contributions plus new ones at each order, it is possible to conjecture the all-loop result from the pattern of explicitly computed terms. We find terms giving rise to $\frac{e^{2 i \gamma \bar\epsilon}}{x^{2-2 i  \bar\epsilon }}$ which is consistent with
$$
    \frac{\Gamma (1-i \bar\epsilon )}{ \Gamma (1+i \bar\epsilon )}=e^{2 i \gamma \bar\epsilon} e^{i\sum_{n=1}^{\infty}\frac{2  (-1)^n }{2 n+1}{\bar\epsilon}^{2 n+1} \zeta(2 n+1)}\, 	
$$    
    in
    $
    32\pi G M^2
\frac{\Gamma (1-i \bar\epsilon )}{ \Gamma (1+i \bar\epsilon )}
\frac{\mathsf{F}_1^2}{4x^{2 (1-i \bar \epsilon )}}
$ in \eqref{eq:SingSub}.
(We note that exponential factors which involve terms such as $\zeta(3)$ and other transcendental constants begin to appear in the loop diagrams from the fourth post-Minkowskian order, which necessitates the appearance of $\zeta(j)$ in the phase factor.)
A different value of $\mu$ will result in an extra phase upon resummation which does not affect physics but prevents a complete match when comparing with black hole perturbation theory.

In the lab frame, the scattering functions are 
\begin{align}
	f^{\rm nt}_{\rm amp}(\theta)+f^{\rm rm}_{\rm amp}(\theta)&={{\cal M}_{\rm finite}(2e^{-\gamma} \omega,\veps^+_1,\veps_2^{+*})\over 8\pi M},\nn\\
	g^{\rm nt}_{\rm amp}(\theta)+g^{\rm rm}_{\rm amp}(\theta)&={{\cal M}_{\rm finite}(2e^{-\gamma} \omega,\veps^+_1,\veps_2^{-*})\over 8\pi M}, 
\end{align}
where $f^{\rm nt}_{\rm amp}(\theta)$ denotes the long-range Newtonian contribution from the first term in \eqref{eq:SingSub} and $f^{\rm rm}_{\rm amp}(\theta)$ gets contributions from the remaining terms. The same holds for the helicity-reversing $g(\theta)$. 

For the resummed Newtonian part, the integration necessary to project to the spherical harmonic basis is well-defined via analytic continuation in $\bar{\epsilon}$.  An example for $l=2$ is 
\begin{align}\label{eq:fntl2}
&f^{\rm nt (2)}_{\rm amp}{=}{\sqrt{5\pi}\over \omega}\bigg[-i+\big(2 \gamma -\tfrac{25}{12}\big) \bar\eps+i\Big(\tfrac{415}{144}-\tfrac{25}{6}\gamma+2  \gamma ^2\Big) \bar\eps^2\nn\\
&\hspace{1cm}- \Big(\tfrac{2}{3}\zeta(3)-\tfrac{25}{6}\gamma^2+\tfrac{4}{3}\gamma^3+\tfrac{415}{72}\gamma-\tfrac{5845}{1728}\Big)\bar\eps^3\bigg].
\end{align}
 The partial wave projection leads to a post-Minkowskian order mixing effect which is clearly illustrated by this example. The $\zeta(3)$ term, expected in the plane wave amplitude at the fourth post-Minkowskian order, actually contributes at the third post-Minkowskian order in the spherical harmonic basis.  This arises  from the singular integration of one term in resummed Newtonian part of the amplitude in \eqref{eq:SingSub},
$
-\frac{8 i \pi }{3 \omega }\int_{0}^1  dx  x (_{-2}Y_{22}){ \bar\eps^4 \zeta (3) \over x^{2-2 i \bar\eps}}=\frac{4 \sqrt{5 \pi } \bar\eps^3 \zeta (3)}{3 \left(\bar\eps^2-3 i \bar\eps-2\right) \omega }$.
The agreement of this term with black hole perturbation theory at $\bar\epsilon^3$-order thus implies that the corresponding term in the Newtonian part in \eqref{eq:SingSub} is correct at the fourth post-Minkowskian order. \black Similarly, the $\bar\epsilon^0$ term in the spherical harmonic component in \eqref{eq:fntl2} actually originates from the first post-Minkowskian amplitude, which explains why we drop the subtraction of $2$ in the definition of $f_B$.

The remaining terms in eq.~\eqref{eq:SingSub} exhibit softer divergences in the projection integral at each post-Minkowskian order and can therefore be integrated directly.  As a highly nontrivial check, we compare the projected components explicitly up to $l=10$ to $\bar\epsilon^3$-order and find exact agreement between the amplitude and black hole perturbation theory results,
\begin{align}
f^{{\rm rm}(l)}_{\rm amp}+ f^{{\rm nt}(l)}_{\rm amp}= f^{(l)}_{\rm B}, &&
g^{{\rm rm}(l)}_{\rm amp}+ g^{{\rm nt}(l)}_{\rm amp}= g^{(l)}_{\rm B} \, .
\end{align} 
The match for higher values of $l$ should be straightforward and \black we expect this equivalence to continue to hold at fourth post-Minkowskian order directly and at higher post-Minkowskian orders after including all the finite-size effects from either completing the effective action, or equivalently, the four-point tree-level Compton amplitude. In particular, the non-perturbative relations \eqref{eq:NPTRelation} appearing on the black hole perturbation theory side suggest the existence of highly non-trivial non-perturbative relations between the complete helicity-conserving Compton amplitude and helicity-reversing one. 

\section{Conclusion and outlook}
In this letter, we explicitly demonstrate that a formalism that systematically incorporates recoil effects in a non-spinning background extends to obtaining the gravitational Compton amplitude to third post-Minkowskian order. We also employ Weinberg's soft theorem to factorize infrared singularities in the Compton amplitude and regulate the leading forward divergences of the long-range Newtonian contributions through a resummed expression. This framework allows us to establish an explicit mapping between perturbative scattering amplitudes in the post-Minkowskian expansion and non-perturbative expressions formulated in terms of spin-weighted spherical harmonics within the Mano-Suzuki-Takasugi solution of black hole perturbation theory. It seems to be worthwhile to formally prove the proposed deduced form for the term regulating the all-order resummed Newtonian contribution by explicitly working to higher orders in the post-Minkowskian expansion.

A natural extension of this work is to incorporate finite-size effects---such as spin, tidal responses, and absorption---into gravitational Compton amplitudes, particularly in light of current astrophysical observations. Because the resummed Newtonian contribution systematically handles post-Minkowskian order mixing, the bridge between formalisms established here indicates that black hole perturbation theory might constrain tree-level Compton amplitudes directly at arbitrary spin order in a way that would enable a bypassing of the conventional separation between near- and far-zone dynamics.

The direct observables in the ringdown stage are the quasi-normal modes~\cite{Vishveshwara:1970zz,Chandrasekhar:1975zza,Leaver:1985ax,Konoplya:2003ii,Berti:2009kk}, which are naturally obtained from computations within black hole perturbation theory. A central open question is how the discrete quasi-normal-mode frequencies and their associated amplitudes arise from post-Minkowskian scattering amplitudes. Our work suggests that such an extraction can be achieved by partially resumming plane-wave amplitudes and systematically mapping them onto partial-wave observables, thereby providing a new pathway to derive ringdown physics from perturbative scattering, which in turn could be used to characterize the black hole spectrum from an action principle. Moreover, generalizing the framework to higher-multiplicity amplitudes is of considerable interest, as it may provide access to nonlinear features of the quasi-normal-mode spectrum~\cite{Lagos:2022otp,Mitman:2022qdl,Kehagias:2023ctr,Cardoso:2026llh,Bourg:2024jme} and thus further deepen the connection between scattering amplitudes and black hole dynamics. We leave these and other questions to future research.

\textit{Note}: During the preparation of our manuscript, we became aware of \cite{Bautista:2026qse} and \cite{Ivanov:2026icp}, which have a partial overlap with us.
\begin{acknowledgements}
{\bf Acknowledgements}
We thank P. L. Ortega for discussions and for collaboration during the initial phase of this work. We also thank K. Haddad, M. Driesse, G. U. Jakobsen, and Y. F. Bautista for coordinating during submission, technical discussions, and comparisons of the results and the master integrals. We also thank M. Ivanov, Y.-Z. Li, J. Parra-Martinez, and Z. Zhou for, through the authors of \cite{Haddad:2026}, coordinating the release of their paper with ours. For discussions on leading singularity analysis, we thank K. Stoldt. We also thank V. Cardoso and J. Redondo-Yuste for useful discussions on black hole perturbation theory and, in particular, the concept of isospectrality of the quasi-normal mode spectrum. We also thank H. Johansson for discussions. The work of N.E.J.B.-B., G.C., and N.S. was supported in part by DFF grant 1026-00077B; The Center of Gravity is a Center of Excellence funded by the Danish National Research Foundation under grant no. DNRF184; and in part by VILLUM Foundation (grant no. VIL37766). The work of C.J.E was funded by the Deutsche Forschungsgemeinschaft (DFG, German Research Foundation), Projektnummer 417533893/GRK2575 `Rethinking Quantum Field Theory'. G.C. has also received funding from the European Union Horizon 2020 research and innovation program under the Marie Sklodowska-Curie grant agreement No. 847523, INTERACTIONS.
\end{acknowledgements}

\bibliographystyle{apsrev4-1}

\bibliography{KinematicAlgebra2}
\newpage

\newpage
\appendix
\onecolumngrid
\begin{center}
    \textbf{\large Supplemental Material -- From the Gravitational Compton Amplitude to Black Hole Perturbation Theory}
\end{center}
In this supplemental material, we present the details of the Compton amplitude computation and derivation of the cross section. We then briefly introduce the method for obtaining the canonical basis and provide explicit expressions for all master integrals. Finally, we present the helicity-conserving and helicity-reversing amplitudes in the lab frame. We also include a detailed comparison between the post-Minkowskian expanded amplitude and the amplitude derived from black hole perturbation theory.    \\ ~ \\
\twocolumngrid

\section{Results for the amplitude and cross section}
From \cite{Bjerrum-Bohr:2025bqg}, the finite parts of the first and second order post-Minkowskian Compton amplitudes are
\vspace{-1.2cm}
\begin{align}\label{eq:2PMFinit}
\mathcal{M}_{0}^{\rm 1PM}
&=(32\pi G) M^2 {\mathsf{F}_1^2\over 4x^2}\nn\\
&\nn \\
\mathcal{M}_{0}^{\rm 2PM}
&=
(32\pi G)^2 M^3 {\omega}\nn\\
&
\!\!\!\!\!\!\!\!\!\!\!\times\Bigg[
\mathsf{F}_1^2 \Bigg(
\frac{i \log \left(\frac{4 \omega ^2}{\mu ^2}\right)}
{64 \pi  x^2 }
-\frac{(x (15 x+28)+15) (x-1)^2}
{1024 x (x+1)^2 (2 x^2-1)}
\nn\\
&
\!\!\!\!\!\!\!\!\!\!\!-\frac{i (4 x^2-7)}
{192 \pi (2 x^4-3 x^2+1)}-\frac{i (2 x^4-2 x^2+1) \log (x)}
{32 \pi  x^2 (x^2-1)^2 (2 x^2-1)}
\Bigg)
\nn\\
&
\!\!\!\!\!\!\!\!\!\!\!+{\mathsf{F}_2^2\over 2x^2-1} \Bigg(
\frac{-i (14 x^4-19 x^2+11)}
{1536 \pi  x^2 (x^2-1)^3 }
+\frac{15 x^2+28 x+15}
{4096 (x+1)^4 x}
\nn\\
&\hspace{3em}
+\frac{i (2 x^4-3 x^2+2) \log (x)}
{128 \pi  (x^2-1)^4 }
\Bigg)
\Bigg]\, .
&\nn \\
&\nn \\
\end{align} 
We note that the product $\mathsf{F}_1\mathsf{F}_2$, used in \cite{Bjerrum-Bohr:2025bqg}, is not linearly independent due to the Gram determinant identity, 
\begin{equation}
    \mathsf{F}_1\mathsf{F}_2 = \frac{\mathsf{F}_2^2 + 4x^2(x^2-1)\mathsf{F}_1^2}{2(1-2x^2)}.
\end{equation}
Starting at the second post-Minkowskian order, the amplitude depends on the fiducial mass scale $\mu$ arising from loop integrals. However, the physical observable--the cross section--must be independent of $\mu$. This can be seen explicitly by factoring out the $\mu$-dependent factors by power counting of the $d$-dimensional coupling in the cross section,
\begin{align}
\lvert \mathcal{M} \rvert^2
&= \left(\frac{d\sigma}{d\Omega}\right)_{\rm LO}  + \left( \frac{\tilde\mu}{\omega} \right)^{2\epsilon}
\left[
\left(\frac{d\sigma}{d\Omega}\right)_{\rm NLO}\Big|_{\tilde\mu \to \omega}
+ \mathcal{O}(\epsilon)
\right] \nn\\[5pt]
& + \left( \frac{\tilde\mu}{\omega} \right)^{4\epsilon}
\left[
\left(\frac{d\sigma}{d\Omega}\right)_{\rm NNLO}\Big|_{\tilde\mu \to \omega}
+ \mathcal{O}(\epsilon)
\right]  + \cdots \, ,
\end{align}\\[5pt]
where $\tilde\mu\to \omega$ means $\mu\to \sqrt{4\pi}e^{-\gamma/2}\omega$ according to the definition.  Similarly, $|\mathcal{M}_{\rm finite}|^2$ is also independent of $\mu$, and therefore it is reasonable to treat the scale parameter $\mu$ appearing in $\mathcal{M}_{\rm finite}$ as a free parameter. 

We list the coefficient of the finite part of the amplitude at third post-Minkowskian order as
\begin{widetext}
\begin{align}
    d_1 &= \frac{8 \mathsf{F}_1^2 x^4 \left(x^2-1\right)^2 \left(8 x^8-30 x^6+48 x^4-35 x^2+10\right)-2 \mathsf{F}_2^2 \left(-55 x^2+2 \left(4
   x^8-21 x^6+54 x^4-77 x^2+63\right) x^4+10\right)}{x^8 \left(x^2-1\right)^4 \left(2 x^2-1\right)}, \nn\\
   d_2 &= -\frac{\left(15 x^6-7 x^4-7 x^2+15\right) \left(4 \mathsf{F}_1^2 \left(x^2-1\right)^2-\mathsf{F}_2^2\right)}{8 x \left(x^2-1\right)^4
   \left(2 x^2-1\right)}, \nn\\
   d_3 &= \frac{4 \mathsf{F}_1^2 \left(x^2-1\right)^2 \left(2 x^4-2 x^2+1\right)+\mathsf{F}_2^2 \left(-2 x^4+3 x^2-2\right) x^2}{2 x^2
   \left(x^2-1\right)^4 \left(2 x^2-1\right)}, \hspace{1cm}
   d_4 = \frac{\left(2 x^4-3 x^2+2\right) \left(4 \mathsf{F}_1^2 \left(x^2-1\right)^2-\mathsf{F}_2^2\right)}{2 \left(x^2-1\right)^4
   \left(2 x^2-1\right)}, \nn\\
   d_5 &= \frac{\mathsf{F}_2^2 \left(517 x^{10}-1650 x^8+2725 x^6-2488 x^4+1200 x^2-240\right)-4 \mathsf{F}_1^2 x^4 \left(x^2-1\right)^2
   \left(341 x^6-730 x^4+693 x^2-240\right)}{24 x^6 \left(x^2-1\right)^4 \left(2 x^2-1\right)}, \nn\\
   d_6 &= \frac{8 \mathsf{F}_1^2 x^2 \left(x^2-1\right)^2 \left(4 x^2-7\right)+\mathsf{F}_2^2 \left(14 x^4-19 x^2+11\right)}{6 x^2
   \left(x^2-1\right)^3 \left(2 x^2-1\right)}, \hspace{1cm}
   d_7 = -\frac{(x (15 x+28)+15) \left(4 \mathsf{F}_1^2
   \left(x^2-1\right)^2-\mathsf{F}_2^2\right)}{16 x (x+1)^4 \left(2 x^2-1\right)}, \nn\\
   d_\text{Re} &= \frac{\mathsf{F}_2^2 \left(36 x^{10}-941 x^8+2275 x^6-2636 x^4+1530 x^2-360\right)-4 \mathsf{F}_1^2 x^4 \left(x^2-1\right)^2
   \left(36 x^6-529 x^4+757 x^2-360\right)}{36 x^6 \left(x^2-1\right)^3 \left(2 x^2-1\right)}, \nn\\
   d_\text{Im} &= i\pi\Big[-\frac{(x-1)^2 \left(4 \mathsf{F}_1^2 \left(x^2-1\right)^2 \left(30 x^6+433 x^5+80 x^4-1091 x^3-560 x^2+720 x+480\right) x^4\right)}{24 x^7 \left(x^2-1\right)^4 \left(2 x^2-1\right)} \\
   -&\frac{(x-1)^2\mathsf{F}_2^2 \left(-30 x^{10}-521 x^9+32
   x^8+2103 x^7+632 x^6-3444 x^5-1664 x^4+2560 x^3+1520 x^2-720 x-480\right)}{24 x^7 \left(x^2-1\right)^4 \left(2 x^2-1\right)}\Big]. \notag
\end{align}
\end{widetext}
The coefficients that enter the next-to-next-to-leading-order cross section are
\begin{align}
c_1 &= 16\left(x^2+\frac{2}{x^2}-3\right), ~~~
c_2 = \frac{15 x^6-7 x^4-7 x^2+15}{4 x^3},\nn\\
c_3 &= \frac{53-255 x^{10}+1169 x^8-2015 x^6+1614 x^4-548 x^2}
{18 x^2 (x^2-1)^2},\nn\\
c_4&=\frac{(x-1)^4 \left(15 x^2+28 x+15\right)^2}{256 x^2 (x+1)^4}-\frac{2 \left(2 x^4-3 x^2+2\right)}{3 x^2},\nn\\
c_5 &= \frac{1}{6 x^2 (x^2-1)^3}\Big(88 x^{12}-547 x^{10}+1319 x^8-1630 x^6\nn\\
&+986 x^4
-211 x^2-29\Big),\nn\\
c_6 &= \frac{4 (x^4-x^2+1)^2 (2 x^4-3 x^2+2)}
{x^2 (x^2-1)^4}\,.
\end{align}
We illustrate the crossing section at $\bar\eps=0.8$, which is a typical magnitude in the ring-down stage.  Interestingly, the next-to-next-to-leading-order contribution dominates the differential cross section at wide scattering angles, spanning $30.0^\circ -90.0^\circ$ for $\bar\epsilon=0.8$, indicating the growing importance of higher-order effects in this regime. 
\begin{figure}
\begin{tikzpicture}[scale=0.9]
\begin{axis}[
title={},   axis lines=left,
			xlabel=$\theta$, 
			ylabel= ${\displaystyle d\sigma\over d\Omega}/ (GM)^2$, ylabel style={rotate=0,at={(-0.07,0.7)},anchor=south},
			legend pos=north east,
			ymajorgrids=true,
			grid style=dashed,
            every axis plot/.append style={thick},enlargelimits=0.05]

\addplot[color=blue, mark=star] coordinates {(72,1.65676) (73.5,1.45376) (75,1.28007) (76.5,1.13167) (78,1.00512) (79.5,0.897482) (81,0.806229) (82.5,0.72919) (84,0.664495) (85.5,0.610532) (87,0.565905) (88.5,0.52941) (90,0.5) (91.5,0.476769) (93,0.458927) (94.5,0.44579) (96,0.436759) (97.5,0.431313) (99,0.428997) (100.5,0.429412) (102,0.43221) (103.5,0.437085) (105,0.443768) (106.5,0.452023) (108,0.461641) (109.5,0.472439) (111,0.484255) (112.5,0.496943)};
\addlegendentry{LO}

\addplot[color=brown, mark=square*,mark size=1pt] coordinates {(72,1.63662) (73.5,1.41737) (75,1.22691) (76.5,1.06143) (78,0.917666) (79.5,0.792775) (81,0.684301) (82.5,0.590116) (84,0.50837) (85.5,0.437457) (87,0.375977) (88.5,0.322714) (90,0.276606) (91.5,0.236729) (93,0.202275) (94.5,0.172539) (96,0.146907) (97.5,0.124842) (99,0.105874) (100.5,0.0895937) (102,0.0756439) (103.5,0.063712) (105,0.0535257) (106.5,0.0448472) (108,0.0374693) (109.5,0.0312118) (111,0.0259173) (112.5,0.0214495)};
\addlegendentry{NLO}

\addplot[color=red, mark=triangle*,mark size=1.5pt] 
coordinates {(72,2.96109) (73.5,2.57713) (75,2.24045) (76.5,1.94527) (78,1.68651) (79.5,1.45976) (81,1.26114) (82.5,1.08727) (84,0.935179) (85.5,0.802245) (87,0.68618) (88.5,0.584964) (90,0.496825) (91.5,0.420198) (93,0.353706) (94.5,0.296138) (96,0.246423) (97.5,0.203618) (99,0.166892) (100.5,0.135513) (102,0.108833) (103.5,0.086283) (105,0.06736) (106.5,0.0516211) (108,0.0386758) (109.5,0.0281802) (111,0.019831) (112.5,0.0133612)};
\addlegendentry{NNLO}
\end{axis}
\end{tikzpicture}
\caption{Differential cross section at $\bar\epsilon=0.8$. }
\label{fig:cs3}
\end{figure}
\section{Differential equations and canonical basis}\label{app:DEs}
We use the method of differential equations \cite{Kotikov:1990kg,Bern:1993kr,Remiddi:1997ny,Gehrmann:1999as,Henn:2013pwa} and find a canonical basis \cite{Henn:2013pwa,Henn:2014qga} to solve the master integrals presented in the main text. To determine the canonical basis, we apply the uniform transcendentality methods of \cite{Dlapa:2021qsl} (see also section 7.1.7 of \cite{Weinzierl:2022eaz}), which, in short, involve choosing master integrals with leading singularities that are constant, linearly independent, and have uniform transcendental weight. In the Baikov representation \cite{Baikov:1996cd,Baikov:1996rk,Baikov:2005nv}, using \texttt{BaikovPackage} \cite{Frellesvig:2024ymq} and imposing the conditions discussed above, we find the following master integrals,
\begin{align}
    \mathcal{K}_8
    &=
    2\eps^2x^2K_7
    +
    \eps^2K_{1,1,1,1,1,1,1}^{1,0,0,0},
    \qquad
    \mathcal{K}_7
    =
    2\eps^2x^2K_7,
    \nn\\
    \mathcal{K}_6
    &=
    \eps^2xK_6,
    \qquad
    \mathcal{K}_5
    =
    \eps^2xK_5,
    \qquad
    \mathcal{K}_4
    =
    \frac{(6\eps - 1)\eps}{4}K_4,
    \nn\\
    \mathcal{K}_3
    &=
    \frac{(6\eps - 1)\eps}{8}K_3,
    \qquad
    \mathcal{K}_2
    =
    \frac{(6\eps - 1)\eps}{8}K_2,
    \nn\\
    \mathcal{K}_1
    &=
    \frac{(1 - 2\eps)^2}{8}K_1.
\end{align}
The differential equation satisfied by these integrals is of canonical form,
\begin{equation}\label{eq:canonical-DE}
    \frac{{\rm d} \mathcal{K}_i(x,\eps)}{{\rm d} x} = \eps\Bigg[\frac{\mathcal{M}_{0,ij}}{x} + \frac{\mathcal{M}_{1,ij}}{1+x} + \frac{\mathcal{M}_{-1,ij}}{1-x}\Bigg]\mathcal{K}_j(x,\eps),
\end{equation}
where $\mathcal{M}$ are the constant coefficient matrices. From the canonical differential equation, we see that the alphabet of the integral family, $\{x,1+x,1-x\}$, is purely logarithmic. Hence, the integrals are expressible purely in terms of multiple polylogarithms. After integrating eq. \eqref{eq:canonical-DE} with \texttt{PolyLogTools} \cite{Duhr:2019tlz} and converting from the canonical to the original starting basis of master integrals, we obtain the presented master integrals,
\begin{widetext}
\begin{align}\label{eq:real-masters-expressions}
K_1 &= -\frac{\omega^2}{16\pi^2}
-\frac{\eps\omega^2}{8\pi^2}
\left[2 + i\pi - \log\frac{4\omega^2}{\mu^2}\right]
-\frac{\eps^2\omega^2}{16\pi^2}
\Bigg[
12 + 8i\pi - 9\zeta(2)
-2\left(4+2i\pi-\log\frac{4\omega^2}{\mu^2}\right)
\log\frac{4\omega^2}{\mu^2}
\Bigg]
+ \mathcal{O}(\eps^3),
\\[4pt]
K_2 &= K_3 =
-\frac{1}{64\pi^2\eps}
-\frac{1}{32\pi^2}
\left[3+i\pi-\log\frac{4\omega^2}{\mu^2}\right]
-\frac{\eps}{64\pi^2}
\Bigg[
36 + 12i\pi - 13\zeta(2)
-2\left(6+2i\pi-\log\frac{4\omega^2}{\mu^2}\right)
\log\frac{4\omega^2}{\mu^2}
\Bigg]\nn\\
&+ \mathcal{O}(\eps^2),
\\[4pt]
K_4 &= -\frac{1}{64\pi^2\eps}
-\frac{1}{32\pi^2}
\left[
3 - 2\log x
-\log\frac{4\omega^2}{\mu^2}
\right]
-\frac{\eps}{64\pi^2}
\Bigg[
36 - 7\zeta(2)
-2\left(
6-\log\frac{4\omega^2x^2}{\mu^2}
\right)
\log\frac{4\omega^2x^2}{\mu^2}
\Bigg]
+ \mathcal{O}(\eps^2),
\\[4pt]
K_5 &= K_6 =
-\frac{i}{256\pi x\omega^2\eps}
+\frac{1}{128\pi^2\omega^2 x}
\Big[
G_-(x)
-i\pi
\left(
\log(1+x)
-\log\frac{8x^2\omega^2}{\mu^2}
\right)
\Big]
+ \mathcal{O}(\eps),
\\[4pt]
K_7 &=
\frac{1}{512\pi^2\omega^4x^2\eps^2}
-\frac{\log\frac{4\omega^2x^2}{\mu^2}}
{256\pi^2\omega^4x^2\eps}
-\frac{1}{512\pi^2\omega^4x^2}
\Big[
\zeta(2)
+4i\pi\log(1+x)
+4G_+(x)
-2\log^2\frac{4\omega^2x^2}{\mu^2}
\Big]
+K_7^{(1)}
+\mathcal{O}(\eps^2),
\\[4pt]
K_8 &=
-\frac{1}{2048\pi^2\omega^6x^2\eps^2}
-\frac{2-x^2-\log\frac{4\omega^2x^2}{\mu^2}}
{1024\pi^2\omega^6x^4\eps}
+\frac{1}{4096\pi^2\omega^6x^4}
\Big[
-5 + x(5x-i\pi(3-4x))
\nonumber\\
&
+2\zeta(2)
+8G_+(x)
+8i\pi\log(1+x)
+4\left(
4-x^2-\log\frac{4\omega^2x^2}{\mu^2}
\right)
\log\frac{4\omega^2x^2}{\mu^2}
-8x^2\log\frac{4\omega^2}{\mu^2}
\Big]
+K_8^{(1)}
+\mathcal{O}(\eps^2).\\
K_7^{(1)} &=
\frac{\eps}{1536\pi^2\omega^4x^2}
\Bigg[
C_7
-32\log^3 x
+6\log^2(1+x)\big(-i\pi+\log x\big)-12G(0,-1,-1,x)
-48G(0,0,-1,x)\nn\\
&
-48G(0,0,1,x)
+12G(0,1,1,x)
-12G(1,0,-1,x)
+12G(1,0,1,x)
+12i\pi\Big(G(1,-1,x)-2G_+(x)\Big)
\nonumber\\
&
+6\log(1-x)
\Big(
3\zeta(2)-G(0,0,x)-2i\pi\log2
\Big)
+48\log^2 x\Big(\log(1-x)-\log z\Big)
+6\Big(\zeta(2)+4G_+(x)\Big)\log z
-4\log^3 z
\nonumber\\
&
+12\log x
\Big(
\zeta(2)-G(1,1,x)-2\log^2 z
\Big)
+6\log(1+x)
\Big[
3\zeta(2)
+4\log x(2i\pi+\log x)
+2G_-(x)
+2i\pi\log(2z^2)
\Big]
\Bigg],
\\[6pt]
K_8^{(1)} &=
\frac{1}{12288\pi^2 x^4\omega^6}
\Bigg[
3-2C_7
+24\zeta(2)
+3x\Big(i\pi+6i\pi x-x(11+16\zeta(2))\Big)
+12\log^2(1+x)\big(i\pi-\log x\big)
+64\log^3 x
\nonumber\\
&
+24G(0,-1,-1,x)
+96G(0,0,-1,x)
+96G(0,0,1,x)
-24G(0,1,1,x)
+24G(1,0,-1,x)
-24G(1,0,1,x)+96G_+(x)
\nonumber\\
&
+6\Big[
-4i\pi G(1,-1,x)
+3xG_-(x)
+8(i\pi-x^2)G_+(x)
+2\log(1-x)
\Big(
-3\zeta(2)
+8G(0,0,x)
+i\pi\log4
\Big)
+i\pi x\log8
\Big]
\nonumber\\
&\quad
+30\log z
-6\Big(
5x^2+i\pi x(-3+4x)
+2\zeta(2)+8G_+(x)
\Big)\log z
+24(-2+x^2)\log^2 z
+8\log^3 z
\nonumber\\
&
+48\log^2 x
\Big(
-4+x^2-2\log(1-x)+2\log z
\Big)\nonumber\\
&
-6\log(1+x)
\Big[
16i\pi\log x
+8\log^2 x
+2(3\zeta(2)+2G_-(x))
+i\pi\Big(-16+x(3+8x)+\log16\Big)
+8i\pi\log z
\Big]
\nonumber\\
&
+12\log x
\Big[
5+3i\pi x-2x^2
-2\zeta(2)
+2G(1,1,x)
+4\log z\big(-4+x^2+\log z\big)
\Big]
\Bigg].
\end{align}
\end{widetext}
We employ the shorthand $z = 4\omega^2/\mu^2$ in the above. The functions $G(\ldots,x)$ are multiple polylogarithms (MPLs), and their definition and properties can be found in, e.g., \cite{Duhr:2019tlz}. We note that all the higher MPLs appearing in $K_7^{(1)}$ and $K_8^{(1)}$ cancel in the amplitude, except for the functions $G_\pm(x)$, whose definition can be found in the main text. Additionally, the $C_7$ that appears above is an undetermined integration constant that also cancels in the amplitude.

\section{Boundary integral values}
 The gravitational Compton amplitude is regular in the backward-scattering limit ($x\to1$). The apparent pole at $x=1$ appearing in the reduced integrand is spurious, as it originates in the tensor and IBP reduction. One can show this by carrying out the integrand reduction with backward-scattering kinematics imposed from the outset. Doing this, one finds that the amplitude is finite, implying our claim that any intermediate divergences one may encounter as $x\to 1$ must be spurious.

By contrast, the forward singularity, proportional to $1/x^2={4\omega^2\over |q|^2}$, is physical and originates from the universal long-range Newtonian gravitational potential $1/r$. A stronger forward singularity would correspond to a long-range interaction incompatible with Einstein gravity. For example, a $1/x^3$ singularity would generate a logarithmic potential, while a $1/x^4$ singularity would correspond to a linearly growing potential. Such behaviors are absent in both Newtonian gravity and general relativity. Therefore, the forward behavior of the amplitude is uniquely fixed to be at most $1/x^2$.

The forward and backward limits thus probe two independent physical properties of the amplitude: the universal long-range gravitational interaction and the regularity of the backward-scattering kinematics, respectively. Together they provide  constraints on  the boundary values,  \black
\begin{align}
\mathcal{K}_{0,7}^{(0)}
&= -\frac{3}{4}\mathcal{K}_{0,1}^{(0)}
-\mathcal{K}_{0,2}^{(0)}, \quad
\mathcal{K}_{0,8}^{(0)}
= \frac{1}{4}\mathcal{K}_{0,1}^{(0)}, \nn\\
\mathcal{K}_{0,7}^{(1)}
&= \frac{1}{4}\left(
-3\mathcal{K}_{0,1}^{(1)}
-4\mathcal{K}_{0,2}^{(1)}
+8\mathcal{K}_{0,5}^{(1)}
\right), \quad 
\mathcal{K}_{0,8}^{(1)}
= \frac{1}{4}\mathcal{K}_{0,1}^{(1)}, \nn\\
\mathcal{K}_{0,7}^{(2)}
&= -\frac{3}{4}\mathcal{K}_{0,1}^{(2)}
+\pi^2\left(
\mathcal{K}_{0,2}^{(0)}
-\frac{1}{4}\mathcal{K}_{0,1}^{(0)}
\right)
-\mathcal{K}_{0,2}^{(2)}
+2\mathcal{K}_{0,5}^{(2)}\nn\\
&
+\log(16)\mathcal{K}_{0,5}^{(1)}, \nn\\
\mathcal{K}_{0,8}^{(2)}
&= \frac{1}{12}\Big(
3\mathcal{K}_{0,1}^{(2)}
-\pi^2\left(
\mathcal{K}_{0,1}^{(0)}
-4\mathcal{K}_{0,2}^{(0)}
\right)\nn\\
&+8\mathcal{K}_{0,2}^{(2)}
-4\mathcal{K}_{0,4}^{(2)}
+8\mathcal{K}_{0,5}^{(2)}
+16\log(2)\mathcal{K}_{0,5}^{(1)}
\Big).
\end{align}
We denote the boundary values of the canonical master integrals by $\mathcal{K}_{0,j}^{(r)}$, where $r=0,1,2$ labels the order in the $\epsilon$ expansion and $j$ labels the master integrals.
 One can see the boundary master integral values for the top topology can be constrained by those of lower topologies, the latter being easy to calculate directly. \black We present their values up to $\mathcal{O}(\epsilon^2)$, 
\begin{align}
\mathcal{K}_{0,1}^{(0)} &= -\frac{1}{128\pi^2},
\qquad
\mathcal{K}_{0,2}^{(0)} =\mathcal{K}_{0,3}^{(0)} = \frac{1}{512\pi^2}, \nn\\
\mathcal{K}_{0,4}^{(0)} &= \frac{1}{256\pi^2},
\qquad
\mathcal{K}_{0,5}^{(0)} =\mathcal{K}_{0,6}^{(0)} = 0, \nn\\
\mathcal{K}_{0,1}^{(1)} &= \frac{\gamma-i\pi-\log(\pi)}{64\pi^2}, 
\quad
\mathcal{K}_{0,2}^{(1)} = \mathcal{K}_{0,3}^{(1)} =\frac{-\gamma+i\pi+\log(\pi)}{256\pi^2}, \nn\\
\mathcal{K}_{0,4}^{(1)} &= -\frac{\gamma-\log(\pi)}{128\pi^2}, 
\qquad
\mathcal{K}_{0,5}^{(1)} = 
\mathcal{K}_{0,6}^{(1)} = -\frac{i}{256\pi}, \nn\\
\mathcal{K}_{0,1}^{(2)}
&=
\frac{\left(2i\gamma+\pi-2i\log(\pi)\right)
\left(2i\gamma+3\pi-2i\log(\pi)\right)}
{256\pi^2}, \nn\\
\mathcal{K}_{0,2}^{(2)}&=\mathcal{K}_{0,3}^{(2)}
=
-\frac{\pi ^2-12 (-\gamma +i \pi +\log (\pi ))^2}{3072 \pi ^2},  \nn\\
\mathcal{K}_{0,4}^{(2)}
&=
-\frac{
-12\gamma^2
+7\pi^2
-12\log^2(\pi)
+24\gamma\log(\pi)}
{1536\pi^2}, \nn\\
\mathcal{K}_{0,5}^{(2)}
&=
\mathcal{K}_{0,6}^{(2)}
=
\frac{i\left(\gamma+\log\left(\frac{2}{\pi}\right)\right)}
{128\pi}.
\end{align}
\section{Boundary integration}\label{app:Boundaries}
In this appendix, we provide a discussion of the integration regions of the third-order post-Minkowskian integral family and evaluate the boundary constants necessary to determine the master integrals that are the particular solution of the system of differential equations. All the boundary constants have been confirmed to be correct by a direct numerical calculation performed with \texttt{pySecDec} \cite{Borowka:2017idc}, including the ones needed for the double box sector, which we do not compute but rather determine by requiring that all spurious poles cancel in the amplitude.

Being able to evaluate the asymptotic expansion of our integrals in the forward-scattering limit $x\to0$ is of particular interest to us. This can be accomplished using the method of regions \cite{Beneke:1997zp,Smirnov:2002pj,Pak:2010pt}. In principle, we only need this to determine the boundary constant for the box-triangles. But, as the regions are quite simple, we provide a discussion at the level of the integral family. Ignoring numerator factors as they are immaterial for the region analysis and integrating out the delta functions, a generic member of the third-order post-Minkowskian integral family for the Compton amplitude is
\begin{align}\label{eq:region-discussion-integral}
    K &= \int_{\ell_1,\ell_2}\frac{\hat\delta(\ell_1\cdot v)\hat\delta(\ell_2\cdot v)}{\ell_1^{2\nu_3}\ell_2^{2\nu_4}(2\omega xQ-\ell_1-\ell_2)^{2\nu_5}}\nn\\
    &\quad \times \frac{1}{(p_1+\ell_1)^{2\nu_6}(p_1+\ell_1+\ell_2)^{2\nu_7}}
\end{align}
where the $\iO$ prescription has been suppressed since it is immaterial for this discussion, and where we have additionally exhibited all dependence on $x$ by defining
\begin{equation}
    Q = \frac{q}{2\omega x}, \qquad Q^2 = -1,
\end{equation}
which uses $|q| = 2\omega x$. The method of regions instructs us to scale the components of the loop momenta with all possible powers of $x$. Of course, the delta functions force the temporal components to vanish, $v\cdot \ell_i = 0$, so we scale the spatial components as
\begin{equation}
    \bell_i \sim x^{p_i},
\end{equation}
and examine which scalings lead to non-scaleless integrals after Taylor expansion of the integrand. We find that two such scalings exist for the integral in \eqref{eq:region-discussion-integral}:
\begin{subequations}
\begin{align}
    \bell_i &\sim x^0 &\text{(normal region)}, \\
    \bell_i &\sim x^1 &\text{(collinear region)}.
\end{align}
\end{subequations}
The leading contribution to the integral in eq. \eqref{eq:region-discussion-integral} in the normal region is
\begin{align}
    K\big\vert_\text{(nor)} &= \int_{\ell_1,\ell_2}\frac{\hat\delta(\ell_1\cdot v)\hat\delta(\ell_2\cdot v)}{\ell_1^{2\nu_3}\ell_2^{2\nu_4}(\ell_1+\ell_2)^{2\nu_5}}\nn\\
    &\quad \times {1\over (p_1+\ell_1)^{2\nu_6}(p_1+\ell_1+\ell_2)^{2\nu_7}}
\end{align}
and clearly scales as $x^0$. The leading contribution to the integral in the collinear region is
\begin{align}\label{eq:generic-collinear-region}
    K\big\vert_\text{(col)} &= \int_{\ell_1,\ell_2}\frac{\hat\delta(\ell_1\cdot v)\hat\delta(\ell_2\cdot v)}{\ell_1^{2\nu_3}\ell_2^{2\nu_4}(2\omega Q-\ell_1-\ell_2)^{2\nu_5}}\nn\\
    &\quad\times\frac{x^{2(3-\nu_3-\nu_4-\nu_5)-\nu_6-\nu_7-4\eps}}{[2p_1\cdot\ell_1]^{\nu_6}[2p_1\cdot(\ell_1+\ell_2)]^{\nu_7}}
\end{align}
Thus, as long as $2(\nu_3+\nu_4+\nu_5)+\nu_6+\nu_7 > 6$, the dominant contribution to the integral in the forward limit will come from the collinear region. We note that, in the region expansion, we have $p_1\cdot q = 0 + \mathcal{O}(x^2)$.

The double bubble and the active double triangles do not depend on $x$. As such, no information about them is gained from the differential equation, and we must hence evaluate them directly. Fortunately, this is quite simple. As this is also the case for the potential double triangle, we evaluate it directly. The box-triangles are dominated by the leading contribution in the collinear region, cf. the preceding section. Therefore, we evaluate them in this region to obtain the boundary constants. As explained in the main text, we refrain from explicitly computing the boundary constants for the integrals in the top sector, instead taking advantage of the physical constraints that the amplitude must obey---namely, the cancellation of all spurious poles.

The double bubble factorizes,
\begin{align}
    K_1 &= \tilde\mu^{4\eps}\int_{\ell_1,\ell_2}\frac{\hat\delta(\ell_1\cdot v)\hat\delta(\ell_2\cdot v)}{[(p_1+\ell_1)^2 + \I0^+][(p_1+\ell_1+\ell_2)^2 + \I0^+]} \nn\\
    &= \tilde\mu^{4\eps}\bigg[\int_{\bell}\frac{1}{\bell^2 - \omega^2 - \I0^+}\bigg]^2 \nn\\
    &= -\frac{e^{-2\I\pi\eps}\Gamma\big(\eps - \frac12\big)^2\tilde\mu^{4\eps}\omega^{2-4\eps}}{(4\pi)^{3-2\eps}},
\end{align}
Next is the potential double triangle,
\begin{equation}
    K_4 = \tilde\mu^{4\eps}\int_{\ell_1,\ell_2}\frac{\hat\delta(\ell_1\cdot v)\hat\delta(\ell_2\cdot v)}{\ell_1^2\ell_2^2(q-\ell_1-\ell_2)^2}.
\end{equation}
As this integral is a three-fold convolution in momentum space, it is a product in position space. Hence, by expanding each propagator in plane waves using
\begin{equation}\label{eq:propagatorFourier}
    \frac{1}{\bell^2} = \frac{2^{d-3}\Gamma\big(\frac{d-3}{2}\big)}{(4\pi)^\frac{d-1}{2}}\int\d^{d-1}\mathbf{x}\,\frac{\euler^{-\I\bell\cdot\mathbf{x}}}{|\mathbf{x}|^{d-3}},
\end{equation}
we obtain the representation
\begin{equation}
    K_4 = \tilde\mu^{4\eps}\frac{8^{d-3}\Gamma\big(\frac{d-3}{2}\big)^3}{(4\pi)^\frac{3(d-1)}{2}}\int\d^{d-1}\mathbf{x}\,\frac{\euler^{-\I\mathbf{q}\cdot\mathbf{x}}}{|\mathbf{x}|^{3(d-3)}},
\end{equation}
which is a standard Fourier integral leading to
\begin{align}
    K_4 &= \frac{\Gamma\big(\frac{1}{2}-\eps\big)^3\Gamma(2\eps)}{(4\pi)^{3-2\eps}\Gamma\big(\frac{3(1-2\eps)}{2}\big)}\bigg(\frac{\tilde\mu^2}{4\omega^2x^2}\bigg)^{2\eps}.
\end{align}
The last integrals we must evaluate exactly are the active double triangles,
\begin{subequations}
\begin{align}
    K_2 &= \tilde\mu^{4\eps}\int_{\ell_1,\ell_2}\frac{\hat\delta(\ell_1\cdot v)\hat\delta(\ell_2\cdot v)}{\ell_1^2\ell_2^2[(p_1+\ell_1+\ell_2)^2 + \I0^+]}, \\
    K_3 &= \tilde\mu^{4\eps}\int_{\ell_1,\ell_2}\frac{\hat\delta(\ell_1\cdot v)\hat\delta(\ell_2\cdot v)}{\ell_2^2(q-\ell_1-\ell_2)^2[(p_1+\ell_1)^2 + \I0^+]}.
\end{align}
\end{subequations}
They are equal, so we only evaluate $K_{1,1,0,1,1,1,0}^{0,0,0,0}$. This can be done loop-by-loop, employing the standard formulae,
\begin{subequations}
\begin{align}   &\int_\ell\frac{\hat\delta(\ell\cdot v)}{\ell^2(k - \ell)^2} = \frac{\Gamma\big(\frac12-\eps\big)^2\Gamma\big(\frac12+\eps\big)}{(4\pi)^{\frac32-\eps}\Gamma(1-2\eps)}\frac{1}{(-k^2)^{\frac12+\eps}}, \label{eq:lbl1}\\    &\int_\ell\frac{\hat\delta(\ell\cdot v)}{(q-\ell)^{2\alpha}[(p_1+\ell)^2 + \iO]} \nn\\
&= \frac{\euler^{\I\pi(2\alpha-\eps)}\Gamma(2-2\alpha-2\eps)\Gamma\big(\alpha+\eps-\frac12\big)\omega^{1-2\alpha-2\eps}}{(4\pi)^{\frac32-\eps}\Gamma(2-\alpha-2\eps)}, \label{eq:lbl2}
\end{align}
\end{subequations}
where the first formula is valid for $k\cdot v = 0$ and the second makes use of the on-shell identity $p_1\cdot q = -q^2/2$. We obtain
\begin{align}
    K_3 &= -\frac{\euler^{2\I\pi\eps}\Gamma(2\eps)\Gamma\big(\frac12-2\eps\big)\Gamma\big(\frac12-\eps\big){\rm sec}\pi\eps}{64\,\Gamma\big(\frac32-3\eps\big)}.
\end{align}
The two box-triangles, 
\begin{subequations}
\begin{align}
    K_5 &= \tilde\mu^{4\eps}\int_{\ell_1,\ell_2}\frac{\hat\delta(\ell_1\cdot v)\hat\delta(\ell_2\cdot v)}{\ell_1^2\ell_2^2(q-\ell_1-\ell_2)^2(p_1+\ell_1+\ell_2)^2}, \\
    K_6 &= \tilde\mu^{4\eps}\int_{\ell_1,\ell_2}\frac{\hat\delta(\ell_1\cdot v)\hat\delta(\ell_2\cdot v)}{\ell_1^2\ell_2^2(q-\ell_1-\ell_2)^2(p_1+\ell_1)^2},
\end{align}
\end{subequations}
are equal, just like the two double triangles. We evaluate the leading contribution to $K_6$ in the collinear region (see eq. \eqref{eq:generic-collinear-region}),
\begin{equation}
    K_6^{0,0,0,0}\big\vert_\text{(col)} = \int_{\ell_1,\ell_2}\frac{x^{-1-4\eps}\hat\delta(\ell_1\cdot v)\hat\delta(\ell_2\cdot v)}{\ell_1^2\ell_2^2(2\omega Q-\ell_1-\ell_2)^2(2p_1\cdot\ell_1+\iO)}.
\end{equation}
One now first integrates out $\ell_2$ using eq. \eqref{eq:lbl1}, after which one can finish the job by using the formula
\begin{align}\label{eq:lbl3}
&\int_\ell\frac{\hat\delta(\ell\cdot v)}{\ell^2(q - \ell)^{2\alpha}[p_1\cdot \ell + \iO]} \nn\\
&= \frac{\I\euler^{\I\pi\alpha}(4\pi)^\eps\Gamma(-\eps)\csc\big(\pi(\alpha+\eps)\big)}{8\Gamma(\alpha)\Gamma(1-\alpha-2\eps)(-q^2)^{\alpha+\eps}\omega} 
    \qquad \text{($p_1\cdot q = 0$)},
\end{align}
which can be derived by introducing Feynman parameters. This procedure yields the result
\begin{align}
    K_6^{0,0,0,0}\big\vert_\text{(col)} &= -\frac{\I\Gamma\big(\frac12-\eps\big){\rm sec}{2\pi\eps}\,\tilde\mu^{4\eps}}{128(2\pi)^{1-2\eps}\eps\Gamma\big(\frac12-3\eps\big)x^{1+4\eps}\omega^{2+4\eps}},
\end{align}
from which one readily determines the requisite boundary constant.

\section{Black hole perturbation theory comparison details}
We present explicitly the remainder parts and the Newtonian contributions of the helicity-conserving and helicity-reversing functions in the lab frame. 
The remainder parts are given in a post-Minkowskian expansion, while the Newtonian contributions are expressed in a resummed form. 
The resummed expression for the helicity-conserving function is necessary to regulate the forward divergence. 
We then project the resummed function onto spin-weighted spherical harmonics, perform a Taylor expansion, and compare the results with those from black hole perturbation theory.
\begin{align}
&f^{\rm rm}_{\rm amp}= \bar\epsilon^2 \Bigg(\frac{i \left(11 x^4-19 x^2+14\right)}{12 \left(1-x^2\right) \omega }+\frac{i \left(2 x^4-3 x^2+2\right) \log (x)}{\left(x^2-1\right)^2 \omega }\nn\\
&+\frac{\pi  (x (15 x+28)+15) (x-1)^2}{32 x (x+1)^2 \omega }\Bigg)\nn\\
&+\bar\epsilon^3\Bigg(\frac{(G_-(x)+i \pi  \log (4x)) \left(15 x^6-7 x^4-7 x^2+15\right)}{16 x \left(x^2-1\right)^2 \omega }\nn\\
&+\frac{(G_+(x)+\log ^2(x)+2 \gamma  \log (x)+\frac{\pi ^2}{6}) \left(-2 x^4+3 x^2-2\right)}{\left(x^2-1\right)^2 \omega }\nn\\
&+\frac{i \pi  \left(x \left(x \left(44 x^2-26 x-51\right)+16\right)+29\right)}{48 (x+1)^2 \omega }\nn\\
&+\frac{\gamma  \left(11 x^4-19 x^2+14\right)}{6 \left(x^2-1\right) \omega }+\frac{170 x^4-127 x^2+53}{72 \omega -72 x^2 \omega }\nn\\
&+\frac{\left(88 x^6-75 x^4+78 x^2-27\right) \log (x)}{24 \left(x^2-1\right)^2 \omega }\nn\\
&+\frac{\gamma  i \pi  (x (15 x+28)+15) (x-1)^2}{16 x (x+1)^2 \omega }\nn\\
&-\frac{i \pi  (x+1)^2 (x (15 x-28)+15) \log (x+1)}{16 (x-1)^2 x \omega }\Bigg)\nn\\
&+\mathcal{O}(\bar\eps^4)\, ,\nn\\
&g^{\rm rm}_{\rm amp}=-\bar\eps^2\frac{11 i  x^2}{12 \omega }+\bar\eps^3 \Bigg(\frac{11 (\gamma+2 \log (x)+\frac{i \pi }{2}) x^2}{6 \omega }\nn\\
&+\frac{(G_+(x)+i \pi  \log (x+1)) \left(6 x^4-15 x^2+10\right)}{x^4 \omega }\nn\\
&-\frac{10 \left(x^2-1\right) \log (x)}{x^2 \omega }-\frac{103 x^4-225 x^2+180}{36 x^2 \omega }\nn\\
&-\frac{i \pi  \left(9 x^4+15 x^3-35 x^2-15 x+30\right)}{3 x^3 \omega }\Bigg)+\mathcal{O}(\bar\eps^4) \, .\nn\\
&f^{\rm nt}_{\rm amp}(\theta)= G M \left(1- x^2\right)^2 x^{-2+2 i \bar\epsilon } \frac{\Gamma (1-i \bar\epsilon )}{\Gamma (1+i \bar\epsilon )},\nn\\
&g^{\rm nt}_{\rm amp}(\theta)=  G M x^{2+2 i \bar\epsilon } \frac{\Gamma (1-i \bar\epsilon )}{\Gamma (1+i \bar\epsilon )}\, .
\end{align}
We compare $f$ and $g$ for all $l \leq 10$ and find agreement. 
Below, we present the results of projecting the $f$ functions onto the modes $l = 2, 3, 4$. 
The corresponding projected values of the $g$ functions are similar, due to the isospectrality of quasi-normal modes (see Eq.~(23) in the main text). 
Note that we omit the overall factor $\sqrt{(2l+1)\pi}/\omega$. 
We also show the reconstruction of $f^{\rm rm}$ and $g$ from the summation over spin-weighted spherical harmonics in fig.~\ref{fig:gmode}. \\[5pt]
\begin{widetext}
\begin{align}
f^{\rm rm (2)}_{\rm amp}&=\frac{\left(-28805+12 \gamma  (1715+1284 i \pi )-16050 i \pi +2568 \pi ^2\right) \bar\eps^3}{15120  }+\frac{(1284 \pi -1715 i) \bar\eps^2}{2520  },\nn\\
f^{\rm nt (2)}_{\rm amp}&=\frac{\bar\eps^3 \left(-1152 \zeta (3)-2304 \gamma ^3+7200 \gamma ^2-9960 \gamma +5845\right)}{1728  }+\frac{\left(415-600 \gamma +288 \gamma ^2\right) i \bar\eps^2}{144  }+\frac{(24 \gamma -25) \bar\eps}{12 }-i,\nn\\
f_{B}^{(2)}&=\frac{\bar\eps^3 \left(-13440 \zeta (3)+29785-88760 \gamma +84000 \gamma ^2-26880 \gamma ^3-21400 i \pi +20544 i \gamma  \pi +3424 \pi ^2\right)}{20160  }\nn\\
&+\frac{i \left(11095-21000 \gamma +10080 \gamma ^2-2568 i \pi \right) \bar\eps^2}{5040  }+\frac{(24 \gamma -25) \bar\eps}{12  }-i\nn\\[0.4cm]
f^{\rm rm (3)}_{\rm amp}&=\frac{\left(-106897+60 \gamma  (581+780 i \pi )-76830 i \pi +7800 \pi ^2\right) \bar\eps^3}{75600  }+\frac{(780 \pi -581 i) \bar\eps^2}{2520  },\nn\\
f^{\rm nt (3)}_{\rm amp}&=\frac{\bar\eps^3 \left(-144000 \zeta (3)-288000 \gamma ^3+1418400 \gamma ^2-2428680 \gamma +1595993\right)}{216000 }\nn\\
&+\frac{\left(20239-23640 \gamma +7200 \gamma ^2\right) i \bar\eps^2}{3600  }+\frac{(120 \gamma -197) \bar\eps}{60  }-i,\nn\\
f^{(3)}_{\rm B}&=\frac{\bar\eps^3 \left(-336000 \zeta (3)+3011337+3309600 \gamma ^2-672000 \gamma ^3-512200 i \pi +52000 \pi ^2+520 i \gamma  (600 \pi +10451 i)\right)}{504000  }\nn\\
&+\frac{i \left(135863-165480 \gamma +50400 \gamma ^2-7800 i \pi \right) \bar\eps^2}{25200  }+\frac{(120 \gamma -197) \bar\eps}{60  }-i\nn\\
f^{\rm rm (4)}_{\rm amp}&=\left(\gamma  \left(\frac{217}{900}+\frac{1571 i \pi }{3465}\right)+\frac{-12938849-11169810 i \pi +942600 \pi ^2}{12474000}\right) \bar\eps^3+\left(\frac{1571 \pi }{6930}-\frac{217 i}{1800}\right) \bar\eps^2,\nn\\
f^{\rm nt (4)}_{\rm amp}&=\bar\eps^3 \left(-\frac{2 \zeta (3)}{3}+\frac{79 \gamma ^2}{10}-\frac{4 \gamma ^3}{3}-\frac{28519 \gamma }{1800}+\frac{829451}{72000}\right)+\left(\frac{28519 i}{3600}-\frac{79 i \gamma }{10}+2 i \gamma ^2\right) \bar\eps^2+\left(2 \gamma -\frac{79}{20}\right) \bar\eps-i,\nn\\
f^{\rm (4)}_{\rm B}&=\bar\eps^3 \left(-\frac{2 \zeta (3)}{3}+\gamma  \left(-\frac{5617}{360}+\frac{1571 i \pi }{3465}\right)+\frac{79 \gamma ^2}{10}+\frac{1571 \pi ^2}{20790}-\frac{4 \gamma ^3}{3}-\frac{124109 i \pi }{138600}+\frac{6792911}{648000}\right)\nn\\
&+\left(\frac{5617 i}{720}-\frac{79 i \gamma }{10}+2 i \gamma ^2+\frac{1571 \pi }{6930}\right) \bar\eps^2+\left(2 \gamma -\frac{79}{20}\right) \bar\eps-i \, .
\end{align}
\begin{figure}
\begin{tikzpicture}[scale=0.9]
\begin{axis}[
title={},   axis lines=left,
			xlabel=$\theta$, 
			ylabel= $f^{\rm rm }$, ylabel style={rotate=-90,at={(-0.07,0.95)},anchor=south},
			legend pos=south east,
			ymajorgrids=true,
			grid style=dashed,
            every axis plot/.append style={thick},enlargelimits=0.05]

\addplot[color=blue, mark=star] coordinates {
(9,-11.9008) (13.5,-5.77349) (18,-3.31091) (22.5,-2.08303)
(27,-1.3894) (31.5,-0.964285) (36,-0.688388) (40.5,-0.50164)
(45,-0.371143) (49.5,-0.277682) (54,-0.209447) (58.5,-0.158876)
(63,-0.120954) (67.5,-0.0922583) (72,-0.070398) (76.5,-0.0536645)
(81,-0.040816) (85.5,-0.030935) (90,-0.0233354) (94.5,-0.0174975)
(99,-0.0130243) (103.5,-0.00960997) (108,-0.00701739)
(112.5,-0.00506188) (117,-0.00359905) (121.5,-0.00251577)
(126,-0.00172334) (130.5,-0.00115224) (135,-0.000748064)
(139.5,-0.000468372) (144,-0.000280197) (148.5,-0.000158076)
(153,-0.0000824956) (157.5,-0.000038648) (162,-0.0000154548)
(166.5,-0.00000479861) (171,-0.000000935121)
(175.5,-0.0000000587315)
};
\addlegendentry{PM}

\addplot[color=brown, mark=square*,mark size=1pt] coordinates {
(9,-2.51933) (13.5,-2.48065) (18,-2.42733) (22.5,-2.36018)
(27,-2.28021) (31.5,-2.18862) (36,-2.08677) (40.5,-1.97613)
(45,-1.85828) (49.5,-1.73487) (54,-1.60758) (58.5,-1.4781)
(63,-1.34807) (67.5,-1.21909) (72,-1.09265) (76.5,-0.970127)
(81,-0.852768) (85.5,-0.741645) (90,-0.637659) (94.5,-0.541524)
(99,-0.45376) (103.5,-0.374693) (108,-0.304455)
(112.5,-0.242999) (117,-0.190103) (121.5,-0.145391)
(126,-0.108352) (130.5,-0.0783599) (135,-0.0547025)
(139.5,-0.036605) (144,-0.0232583) (148.5,-0.0138467)
(153,-0.00757513) (157.5,-0.00369481) (162,-0.00152749)
(166.5,-0.000486803) (171,-0.0000966545)
(175.5,-0.00000605957)
};
\addlegendentry{$l_{\rm max}=2$}

\addplot[color=red, mark=triangle*,mark size=1.5pt] 
coordinates {
(9,-14.3534) (13.5,-12.2023) (18,-9.56943) (22.5,-6.76167)
(27,-4.08523) (31.5,-1.80137) (36,-0.091533) (40.5,0.962651)
(45,1.38189) (49.5,1.27488) (54,0.810707) (58.5,0.183664)
(63,-0.422607) (67.5,-0.865297) (72,-1.06194) (76.5,-0.996691)
(81,-0.713467) (85.5,-0.298596) (90,0.142657) (94.5,0.510438)
(99,0.730634) (103.5,0.767515) (108,0.626949)
(112.5,0.350555) (117,0.00298589) (121.5,-0.344186)
(126,-0.628167) (130.5,-0.806863) (135,-0.864818)
(139.5,-0.812558) (144,-0.680254) (148.5,-0.507859)
(153,-0.334472) (157.5,-0.189605) (162,-0.0882875)
(166.5,-0.0307933) (171,-0.00651343)
(175.5,-0.000423964)
};
\addlegendentry{$l_{\rm max}= 6$}
\addplot[color=gray, mark=diamond*,mark size=1.5pt] 
coordinates {
 (13.5,-15.3721) (18,-6.87189) (22.5,-0.867378)
(27,1.80674) (31.5,1.64605) (36,0.0273059) (40.5,-1.53974)
(45,-2.08479) (49.5,-1.4778) (54,-0.276109) (58.5,0.732901)
(63,1.0149) (67.5,0.545267) (72,-0.259397) (76.5,-0.847655)
(81,-0.873433) (85.5,-0.373754) (90,0.297605) (94.5,0.714716)
(99,0.64182) (103.5,0.158787) (108,-0.411057)
(112.5,-0.711382) (117,-0.566997) (121.5,-0.0762648)
(126,0.462427) (130.5,0.734272) (135,0.58666)
(139.5,0.100709) (144,-0.47495) (148.5,-0.875927)
(153,-0.96288) (157.5,-0.772206) (162,-0.460335)
(166.5,-0.19132) (171,-0.0455303)
(175.5,-0.00317258)
};
\addlegendentry{$l_{\rm max}= 10$}
\end{axis}
\end{tikzpicture}
\begin{tikzpicture}[scale=0.9]
\begin{axis}[
title={},   axis lines=left,
			xlabel=$\theta$, 
			ylabel= $g$, ylabel style={rotate=-90,at={(-0.07,1.0)},anchor=south},
			legend pos=south east,
			ymajorgrids=true,
			grid style=dashed,
            every axis plot/.append style={thick},enlargelimits=0.05]

\addplot[color=blue, mark=star] coordinates {
(9,-0.0140068) (13.5,-0.0267589) (18,-0.0414509) (22.5,-0.0572201)
(27,-0.0733901) (31.5,-0.0894119) (36,-0.104834) (40.5,-0.119286)
(45,-0.132464) (49.5,-0.144127) (54,-0.154088) (58.5,-0.162213)
(63,-0.168411) (67.5,-0.172637) (72,-0.174883) (76.5,-0.175178)
(81,-0.173586) (85.5,-0.170198) (90,-0.165133) (94.5,-0.158533)
(99,-0.15056) (103.5,-0.141392) (108,-0.131223) (112.5,-0.120255)
(117,-0.108696) (121.5,-0.096761) (126,-0.0846627) (130.5,-0.072613)
(135,-0.0608182) (139.5,-0.0494767) (144,-0.0387764) (148.5,-0.0288921)
(153,-0.0199836) (157.5,-0.0121936) (162,-0.00564573) (166.5,-0.000443355)
(171,0.00333183) (175.5,0.00562076)
};
\addlegendentry{PM}

\addplot[color=brown, mark=square*,mark size=1pt] coordinates {
(9,-5.46843e-6) (13.5,-2.75419e-5) (18,-8.6421e-5) (22.5,-0.000209042)
(27,-0.000428579) (31.5,-0.000783408) (36,-0.00131589) (40.5,-0.00207101)
(45,-0.00309491) (49.5,-0.00443337) (54,-0.00613023) (58.5,-0.00822582)
(63,-0.0107555) (67.5,-0.0137482) (72,-0.0172252) (76.5,-0.021199)
(81,-0.0256725) (85.5,-0.0306379) (90,-0.0360769) (94.5,-0.0419601)
(99,-0.0482472) (103.5,-0.054887) (108,-0.0618188) (112.5,-0.0689724)
(117,-0.0762698) (121.5,-0.0836264) (126,-0.0909522) (130.5,-0.0981538)
(135,-0.105136) (139.5,-0.111803) (144,-0.118063) (148.5,-0.123826)
(153,-0.129008) (157.5,-0.133532) (162,-0.137331) (166.5,-0.140348)
(171,-0.142537) (175.5,-0.143863)
};
\addlegendentry{$l_{\rm max}=2$}

\addplot[color=red, mark=triangle*,mark size=1.5pt] 
coordinates {
(9,-0.000421589) (13.5,-0.00204428) (18,-0.00608166) (22.5,-0.0137327)
(27,-0.0258735) (31.5,-0.0427757) (36,-0.0639438) (40.5,-0.0881103)
(45,-0.113395) (49.5,-0.137596) (54,-0.15856) (58.5,-0.174547)
(63,-0.18452) (67.5,-0.188304) (72,-0.186565) (76.5,-0.180642)
(81,-0.172243) (85.5,-0.163083) (90,-0.154546) (94.5,-0.147426)
(99,-0.141813) (103.5,-0.137141) (108,-0.13238) (112.5,-0.126334)
(117,-0.117966) (121.5,-0.106689) (126,-0.0925477) (130.5,-0.076255)
(135,-0.0590783) (139.5,-0.0425937) (144,-0.0283684) (148.5,-0.01764)
(153,-0.0110638) (157.5,-0.00858399) (162,-0.00945831) (166.5,-0.0124303)
(171,-0.0160126) (175.5,-0.0188189)
};
\addlegendentry{$l_{\rm max}= 6$}
\addplot[color=gray, mark=diamond*,mark size=1.5pt] 
coordinates {
(9,-0.00146576) (13.5,-0.0066813) (18,-0.0182336) (22.5,-0.0368802)
(27,-0.0608618) (31.5,-0.0864205) (36,-0.109356) (40.5,-0.126795)
(45,-0.138256) (49.5,-0.145482) (54,-0.151216) (58.5,-0.157597)
(63,-0.165083) (67.5,-0.172393) (72,-0.177432) (76.5,-0.178584)
(81,-0.175613) (85.5,-0.1697) (90,-0.162653) (94.5,-0.155828)
(99,-0.149409) (103.5,-0.142469) (108,-0.133717) (112.5,-0.122455)
(117,-0.10913) (121.5,-0.0951492) (126,-0.0820982) (130.5,-0.0708432)
(135,-0.0610946) (139.5,-0.051702) (144,-0.0415062) (148.5,-0.0302213)
(153,-0.0187933) (157.5,-0.00899176) (162,-0.00245519) (166.5,0.000249513)
(171,-0.0000121474) (175.5,-0.00139584)
};
\addlegendentry{$l_{\rm max}= 10$}
\end{axis}
\end{tikzpicture}
\caption{Left: Reconstruction of $f^{\rm rm}(\theta)$ from summation of the spin-weighted harmonic modes, $\sum_{l=2}^{ l_{\rm max}} f^{\text{rm}(l)} {}_{-2}Y_{l2}(\theta)$. Right: Reconstruction of $g(\theta)$ from summation of the spin-weighted harmonic modes, $\sum_{l=2}^{ l_{\rm max}} g^{(l)}_{B} {}_{-2}Y_{l2}(\pi-\theta)$.}
\label{fig:gmode}
\end{figure}
\end{widetext}
\end{document}